\documentclass[12pt]{JHEP3}
\usepackage{amsmath,epsf,amssymb,latexsym,cite,graphics,setspace,bbm}
\usepackage[matrix,arrow,frame,import,curve,color]{xy}
\usepackage[active]{srcltx}

\DeclareFontFamily{U}{rsf}{}
\DeclareFontShape{U}{rsf}{m}{n}{
  <5> <6> rsfs5 <7> <8> <9> rsfs7 <10-> rsfs10}{}
\DeclareMathAlphabet\Scr{U}{rsf}{m}{n}

\def\C{{\mathbb C}}

\def\P{{\mathbb P}}
\def\R{{\mathbb R}}
\def\Z{{\mathbb Z}}

\def\Aut{\operatorname{Aut}}
\def\End{\operatorname{End}}

\def\deg{\operatorname{deg}}

\def\im{\operatorname{im}}

\def\ri{\operatorname{ri}}

\def\GL{\operatorname{GL}}

\def\GU{\operatorname{U{}}}

\def\p{\partial}

\def\la{\langle}
\def\ra{\rangle}
\def\lad{\langle\!\langle}
\def\rad{\rangle\!\rangle}

\def\ff#1#2{{\textstyle\frac{#1}{#2}}}

\def\cD{{\cal D}}
\def\cE{{\cal E}}
\def\cF{{\cal F}}

\def\cL{{\cal L}}
\def\cM{{\cal M}}

\def\cO{{\cal O}}

\def\cV{{\cal V}}
\def\cW{{\cal W}}

\def\ep{{\epsilon}}

\newcommand\thetab{\overline{\theta}}

\newcommand\rhot{\widetilde{\rho}}

\newcommand\Gammat{\widetilde{\Gamma}}

\newcommand\Mt{\widetilde{M}}

\newcommand\Vt{\widetilde{V}}
\newcommand\Wt{\widetilde{W}}

\def\Ve{{\cV}}
\def\cDb{\overline{\cD}}

\def\Autt{\widetilde{\Aut}}
\def\hk#1{h^{1,1}_{\text{toric}}(#1)}
\def\hc#1{h^{d-2,1}_{\text{poly}}(#1)}

\def\GUR{{\GU(1)_{\text{R}}}}
\def\GUL{{\GU(1)_{\text{L}}}}

\title{(0,2) Deformations of Linear Sigma Models}
\author{Maximilian Kreuzer\\
\normalsize Institute for Theoretical Physics \\
TU-Wien, \\
Wiedner Hauptstr. 8-10 \\
1040 Vienna, Austria \\
Email: \email{kreuzer@hep.itp.tuwien.ac.at}
}
\author {Jock~McOrist\\
\normalsize Department of Applied Mathematics and Theoretical Physics \\
\normalsize Centre for Mathematical Sciences\\
\normalsize Wilberforce Road,
\normalsize Cambridge, CB3 OWA, UK \\
Email:  \email{j.mcorist@damtp.cam.ac.uk}
}
\author{Ilarion V.~Melnikov\\
\normalsize Max-Planck-Institut f\"ur Gravitationsphysik (Albert-Einstein-Institut),\\
\normalsize Am M\"uhlenberg 1, D-14476 Golm, Germany\\
Email:  \email{ilarion@aei.mpg.de}
}
\author{M.~Ronen Plesser\\
\normalsize Center for Geometry and Theoretical Physics, Box 90318\\
\normalsize Duke University, Durham, NC 27708-0318, USA\\
Email: \email{plesser@cgtp.duke.edu}
}

\abstract{ We study (0,2) deformations of a (2,2) supersymmetric gauged linear sigma model for a Calabi-Yau hypersurface in a Fano toric variety.  In the non-linear sigma model these correspond to some of the holomorphic deformations of the tangent bundle on the hypersurface.  Combinatorial formulas are given for the number of these deformations, and we show that these numbers are exchanged by mirror symmetry in a subclass of the models.} 

\preprint{AEI-2010-003 \\ DAMTP-2010-2}

\begin{document}

\section{Introduction}
Mirror symmetry is by now a familiar property of two-dimensional (2,2) superconformal field theories (SCFTs).  In the modern times it is mostly stated, studied, and applied in the context of type II string theories, with only occasional terminology, such as ``Yukawa couplings,'' indicating its origins in the study of heterotic strings.  In that context, the restriction to theories with (2,2) supersymmetry is rather artificial, since most (2,2) heterotic theories have deformations preserving only (0,2) superconformal invariance.  The mirror isomorphism should extend to these deformations, but currently the extension is not understood in any detail.  In this work we will identify a class of (2,2) models and (0,2) deformations where  the number of (0,2) deformations is mirror symmetric, and the (0,2) deformation space admits a simple algebraic description.  In order to discuss these models, we will first present a quick and prejudiced sketch of (2,2) mirror symmetry.  A thorough pedagogical treatment is given in~\cite{Cox:2000vi,Hori:2003ds}.

The basic statement of mirror symmetry is that $d$-dimensional
Calabi-Yau manifolds come in pairs $(M,M^\circ)$, and the pair of
non-linear sigma models (NLSMs) with target spaces $M$, $M^\circ$
define families of isomorphic (2,2) SCFTs.  Near a generic point the
moduli space for the SCFT obtained from $M$ is a product manifold
$\cM_{\text{cK}} (M) \times \cM_{\text{c-x}} (M)$, where
$\cM_{\text{cK}}$ parametrizes the choice of complexified K\"ahler
class, and $\cM_{\text{c-x}}$ labels the choice of complex structure.
$\cM_{\text{cK}}$ and $\cM_{\text{c-x}}$ are special K\"ahler
manifolds of dimensions $h^{1,1}(M)$ and $h^{d-1,1}(M)$, respectively.
Mirror symmetry implies that $\cM_{\text{cK}}(M)$ and
$\cM_{\text{c-x}}(M^\circ)$ are isomorphic as special K\"ahler
manifolds, a statement that goes a long way to finding the explicit
isomorphism and is a crucial ingredient in the remarkable enumerative
predictions of mirror symmetry~\cite{Candelas:1990qd}.

A large class of mirror pairs consists of Calabi-Yau (C-Y)
hypersurfaces in certain Fano toric varieties~\cite{Batyrev:1994hm}.
The three-folds obtained via this construction were classified by
computer search~\cite{Kreuzer:2000xy}, which produced $30108$ pairs of
distinct Hodge numbers.  This construction has been extended to
Calabi-Yau complete intersections in toric
varieties~\cite{Batyrev:1994pg,Batyrev:2007cq}. This much larger set
remains unclassified, and it is unknown whether it is finite.  The
NLSMs for all of these manifolds are distinguished by having a gauged linear
sigma model (GLSM) description~\cite{Witten:1993yc}---they are
realized as IR fixed points of a 
two-dimensional (2,2) supersymmetric abelian gauge theory.
In this context
mirror symmetry can be stated as an equivalence of IR limits of two
comparatively simple UV theories: a GLSM for $M$ and a GLSM for
$M^\circ$.  In what follows, we will restrict attention to the case of
hypersurfaces, but much of what we will discuss in this note should be
extendable to the complete intersection case at the price of
additional combinatorics.

The GLSM perspective is useful for describing the SCFT deformations
that are easily identified with simple terms in the UV action.
General considerations of the renormalization group imply that each of
the deformations of the fixed point must be identifiable with some
GLSM deformation; however, the question is whether an SCFT deformation
corresponds to a deformation of some simple part of the GLSM
Lagrangian, such as a holomorphic superpotential.  This is the case
for the subspace of ``toric'' K\"ahler deformations---elements of
$H^{1,1}(M)$ obtained by pulling back elements of $H^{1,1}(V)$ and
for the subspace of ``polynomial'' complex structure
deformations---deformations of complex structure of $M$ that arise by
deforming the defining hypersurface equation.  As we will review,
these do not give a complete description of the SCFT moduli space.
Remarkably, however, these subspaces are preserved by the mirror
isomorphism: toric deformations of $M$ map to polynomial deformations
of $M^\circ$ and vice versa.  The resulting ``algebraic gauge''
coordinates combined with the monomial divisor mirror
map~\cite{Aspinwall:1993rj} are natural for explicit GLSM
computations~\cite{Morrison:1994fr} and may be used to prove mirror
symmetry---at least at the level of topological theory---without
relying on special K\"ahler coordinates~\cite{Batyrev:2002em,
  Szenes:2004mv,Borisov:2005hs,Karu:2005tr}.

In this work we will be concerned with (0,2) deformations of (2,2) GLSMs.  The study of such theories is of great interest for probing the stringy geometry of the heterotic string and serves as a useful starting point for exploring quantum corrections in more general (0,2) theories.  The (0,2) GLSMs have already played an important role in understanding heterotic compactifications.  The original construction of~\cite{Witten:1993yc} was explored in detail in~\cite{Distler:1993mk}, where it was shown that special points in the moduli space could be used to obtain certain exact results;  in~\cite{Silverstein:1995re,Basu:2003bq,Beasley:2003fx} the GLSM perspective was used to argue for the existence of non-trivial (0,2) deformations; and, more recently, studies of the half-twisted GLSM have led to an improved understanding of deformations of quantum cohomology~\cite{Adams:2003zy,Katz:2004nn,Guffin:2007mp,McOrist:2007kp,McOrist:2008ji}.

Given a (2,2) GLSM for some C-Y $M$ in a Fano toric variety $V$, we
know exactly how to construct the GLSM for the mirror hypersurface,
$M^\circ \subset V^\circ$, and for both of the GLSMs there are natural
deformations of the GLSM Lagrangian that preserve (0,2)
supersymmetry.\footnote{Working in a geometric phase of the GLSM it is
  easy to see that infinitesimally these deformations correspond to
  some of the unobstructed elements of $H^1(\End T_M)$.} These GLSM
deformations are believed to describe some of the (0,2) deformations
of the IR fixed
point~\cite{Silverstein:1995re,Basu:2003bq,Beasley:2003fx}.  Mirror
symmetry, when thought of as an isomorphism of two SCFTs, implies an
isomorphism of their (0,2) deformations as well.  Does this map yield an
isomorphism from one set of (0,2) GLSM deformations to the other?
There is no a priori reason why this should be so, just as there was
no a priori reason for the (2,2) mirror map should respect the
splitting of moduli into toric/non-toric and
polynomial/non-polynomial.  Nevertheless, we find evidence that there
is such a split in a subclass of GLSMs.

Our evidence is much cruder than that for the original
monomial-divisor mirror map~\cite{Aspinwall:1993rj} and does not yield
an explicit isomorphism: in fact, all we verify is that the
numbers of (0,2) deformations agree.  However, this is a necessary
first step in seeking to construct the map---a topic to which we plan
to return in the near future.  We can now state our main results:
\begin{enumerate}
\item Given a (2,2) mirror pair of GLSMs for $(d-1)$-dimensional C-Y
  hypersurfaces $M \subset V$ and $M^\circ \subset V^\circ$, the
  number of (0,2) deformations of the GLSMs is mirror symmetric
  provided that $\dim \Aut V = d$ and $\dim \Aut V^\circ = d$,
  i.e. the automorphism groups of $V$ and $V^\circ$ have minimal
  dimension for $d$-dimensional toric varieties.
\item When either member of the pair fails to meet the condition, the
  numbers are generally not mirror symmetric.  The form of the discrepancy
  suggests that some modification of the GLSM could restore the
  symmetry; however, we have not been able to find a modification that
  leads to mirror symmetric results in all cases.
\end{enumerate}

Our results lead to a number of questions.  First, in the favorable
case where the number of (0,2) deformations is mirror symmetric, what
is the mirror map at the level of the GLSM?  How does this map act on
the half-twisted correlators?  Does this lead to computationally
useful results?  More generally, what is the number of deformations of
the tangent bundle for a Calabi-Yau hypersurface in a toric variety?
We suspect this should be given by a combinatorial formula
generalizing our counting of GLSM parameters, and the correction
terms, akin to those describing non-toric and non-polynomial
deformations, should combine with our results to give mirror symmetric
counts.\footnote{Such a result may perhaps hold only with suitable
  genericity assumptions on the (2,2) parameters. After all, it is
  known that $\dim H^1(\End T_M)$ is in general a discontinuous
  function of complex structure moduli~\cite{Hubsch:1992nu}, and mirror symmetry leads us
  to expect similar discontinuities for suitably tuned K\"ahler
  parameters.}

The rest of the paper is organized as follows: after reviewing some
standard notions of toric geometry in section~\ref{s:toric}, we turn
to the GLSM (2,2) Lagrangian and its (0,2) deformations in
section~\ref{s:action}; in section~\ref{s:combin} we give
combinatorial expressions for the dimension of the space of
deformations, apply them to a number of examples, and prove our basic
results; finally, we discuss the mismatch of (0,2) parameters in
generic mirror pair GLSMs in section~\ref{s:discussion}.  The appendix
describes the parameter space of (0,2) deformations for the GLSM for a
toric variety $V$.

\acknowledgments  It is a pleasure to thank P.~Aspinwall, P.~Candelas, J.~Distler, C.~Haase, B.~Nill, and X.~de la Ossa for useful discussions.  The work of MK is supported by the Austrian Research Funds FWF grants no. I192 and P21239.  JM is supported in part by the EPSRC Postdoctoral Fellowship EP/G051054/1. The work of IVM is supported in part by the German-Israeli Project cooperation (DIP H.52) and the German-Israeli Fund (GIF).
MRP is supported in part by the NSF grant DMS-0606578.

\section{Some Toric Notions} \label{s:toric}
We begin with a review of some toric geometry relevant to the construction of mirror pairs of Calabi-Yau hypersurfaces in toric varieties~\cite{Batyrev:1994hm}.  This material is well-known, and a pedagogical treatment is given in~\cite{Cox:2000vi}.

\vspace{0.5cm}

\subsection{Polytopes}
\begin{figure}[h!]
\[
\begin{xy} <1.0mm,0mm>:
(-30,0)*\xybox{
  (0,0)*{} ="0", (25,0)*{}="pX", (-25,0)*{}="mX",
  (0,25)*{} = "pY", (0,-25)*{} ="mY", (15,20)*{\Delta \subset M_\R},
 (-5, 10)*{\bullet}="v1", (-5, 5)*{\bullet}, (-5,0)*{\bullet}, (-5,-5)*{\bullet}="v2",  
 ( 0,   5)*{\bullet}, (0, 0)*{\bullet}, (0,-5)*{\bullet},
 ( 5,  0)*{\bullet}, (5, -5)*{\bullet}, (10,-5)*{\bullet}="v3",
\ar@{-}|{} "mX"; "pX" 
\ar@{-}|{} "mY"; "pY"
\ar@{-}|{} "v1"; "v2"
\ar@{-}|{} "v2"; "v3"
\ar@{-}|{}"v3"; "v1"
} ,
(30,0)*\xybox{
  (0,0)*{} ="0", (25,0)*{}="pX", (-25,0)*{}="mX",
  (0,25)*{} = "pY", (0,-25)*{} ="mY", (15,20)*{\Delta^\circ \subset N_\R},
  (0,5)*{\bullet}="v1", (5,0)*{\bullet}="v2" , (0,0)*{\bullet} , (-5,-5)*{\bullet}="v3",
\ar@{-}|{} "mX"; "pX" 
\ar@{-}|{} "mY"; "pY"
\ar@{-}|{} "v1"; "v2"
\ar@{-}|{} "v2"; "v3"
\ar@{-}|{}"v3"; "v1"
} 
\end{xy}
\]
\caption{Reflexive polytopes for the elliptic curve in $\P^2$.}
\label{fig:torus}
\end{figure}
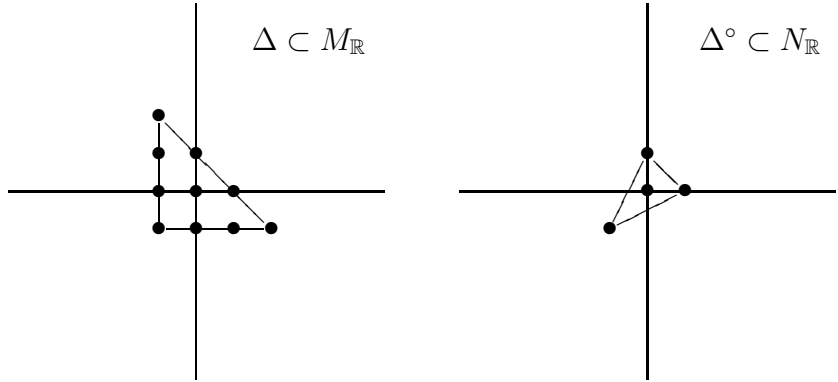

Let $M$ be a $d$-dimensional lattice in $M_{\R} \simeq \R^d$,  $N  \in N_\R \simeq (M_{\R})^{\vee}$ be the dual lattice, and denote the natural
pairing $M_{\R}\times N_{\R} \to \R$ by $\la\cdot,\cdot\ra$.  Let
$\Delta \subset M_\R$ be a lattice polytope (i.e. one whose vertices lie in $M$) containing the origin.  The dual polytope $\Delta^\circ \subset N_{\R}$ is defined by
\begin{equation}
\Delta^\circ = \left\{ y\in N_{\R}~|~ \la x,y \ra \ge -1~ \forall~ x \in \Delta \right\}.
\end{equation}
A familiar example from  $d=2$ is given in figure~\ref{fig:torus}.
The lattice polytope $\Delta$ is said to be reflexive if and only if its polar dual $\Delta^\circ$ is also a lattice polytope.  Since $(\Delta^\circ)^\circ = \Delta$, it follows that $\Delta$ is reflexive if and only if $\Delta^\circ$ is reflexive.  Since $\Delta$ is defined by hypersurfaces that are distance $1$ from the origin, it follows that $\Delta$ must have a unique interior point.

There are two natural ways to relate $\Delta$ to geometry:
\begin{enumerate}
\item $\Delta$ is a Newton polytope for a hypersurface $\{ P = 0\} \subset (\C^\ast)^d$;
\item $\Delta$ defines a complete toric variety $V$ with fan $\Sigma_V \subset N_{\R}$ given by taking cones over faces of $\Delta^\circ$.  In particular, the one-dimensional cones, denoted by $\Sigma_V(1)$, correspond to vertices of $\Delta^\circ$.
\end{enumerate}
In fact, these two notions are nicely related:  the hypersurface $\{P = 0\} \subset (\C^\ast)^d$ has a natural compactification to a subvariety 
$M = \overline{ \{P = 0\} } \subset V. $ When $\Delta$ is reflexive, then $V$ is Fano, and $M \subset V$ is a Calabi-Yau hypersurface with ``suitably mild'' singularities~\cite{Batyrev:1994hm}.

\begin{figure}[h]
\[
\begin{xy} <1.25mm,0mm>:
(-20,0)*\xybox{
  (-7,-5)*{3},(-7,11)*{1},(10,-7)*{2},
  (0,0)*{} ="0", (25,0)*{}="pX", (-25,0)*{}="mX",
  (0,25)*{} = "pY", (0,-25)*{} ="mY", 
 (-5, 10)*{\bullet}="v2", (-10,20)*{} ="L2",
 (-5, 5)*{\bullet}, (-5,0)*{\bullet}, 
 (-5,-5)*{\bullet}="v1", (-10,-10)*{}="L1", 
 ( 0,   5)*{\bullet}, (0, 0)*{\bullet}, (0,-5)*{\bullet},
 ( 5,  0)*{\bullet}, 
 (5, -5)*{\bullet}, 
 (10,-5)*{\bullet}="v3", (20,-10)*{}="L3",
\ar@{-}|{}"0";"L2"
\ar@{-}|{}"0";"L1"
\ar@{-}|{}"0";"L3"
},
(20,0)*\xybox{
 (0,0)*{} ="0", (25,0)*{}="pX", (-25,0)*{}="mX",
  (0,25)*{} = "pY", (0,-25)*{} ="mY", 
 (-5, 10)*{\bullet}="v2", (-10,20)*{} ="L2", (-7,11)*{1},
 (-5, 5)*{\bullet}, (-10,10)*{}="L6", (-7,5)*{7},
 (-5,0)*{\bullet},  (-13,0)*{} = "L7", (-7,1)*{9},
 (-5,-5)*{\bullet}="v1", (-10,-10)*{}="L1", (-7,-5)*{3},
 ( 0,   5)*{\bullet}, (0,13)*{}="L8", (-1,7)*{8},
 (0, 0)*{\bullet}, 
 (0,-5)*{\bullet}, (0,-13)*{}="L9", (-1,-7)*{4},
 ( 5,  0)*{\bullet}, (13,0)*{} = "L5", (6,2)*{6},
 (5, -5)*{\bullet}, (10,-10)*{}="L4", (5,-7)*{5},
 (10,-5)*{\bullet}="v3", (20,-10)*{}="L3", (10,-7)*{2},
\ar@{-}|{}"0";"L2"
\ar@{-}|{}"0";"L1"
\ar@{-}|{}"0";"L3"
\ar@{-}|{}"0";"L4"
\ar@{-}|{}"0";"L5"
\ar@{-}|{}"0";"L6"
\ar@{-}|{}"0";"L7"
\ar@{-}|{}"0";"L8"
\ar@{-}|{}"0";"L9"
}
\end{xy}
\]
\caption{Fan for the mirror elliptic curve (left) and its maximal projective subdivision.}
\label{fig:mirtor}
\end{figure}
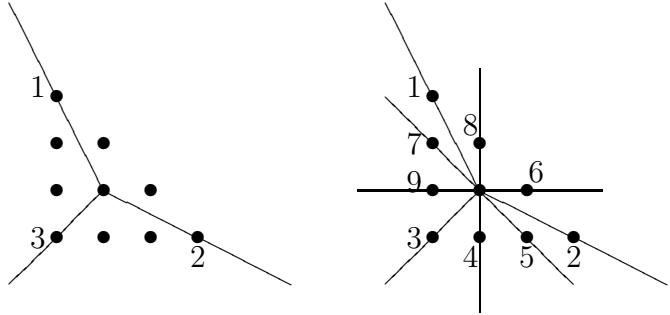

Let us make more precise the notion of  ``suitably mild singularities.''  A toric variety $V$ with fan $\Sigma_V$ is smooth if and only if every full dimensional cone $\sigma \in \Sigma_V$ is simplicial, and its generators furnish a basis for $N$.  In general, the $\Sigma_V$ constructed as above will lead to a singular variety $V$.  For example, exchanging the roles of $\Delta,\Delta^\circ$ in the example of figure~\ref{fig:torus} yields the fan shown in the left half of figure~\ref{fig:mirtor}.
The resulting toric variety is $\C\P^2/\Z_3$--a space with three singular points.  As shown in the figure, the fan may be refined by introducing six additional one-dimensional cones, leading to a fan $\Sigma_{\Vt}$ for a smooth projective variety $\Vt$.   This construction gives a toric crepant resolution of singularities $\Vt \to V$.

Although any toric variety has a toric resolution of singularities obtained by refining the fan, in general these will fail to be crepant.  However, for the purposes of constructing smooth Calabi-Yau hypersurfaces, there is a toric crepant resolution of singularities that is sufficiently smooth.  Namely,  any fan $\Sigma_V$ constructed from $\Delta^\circ$ may be refined to a fan $\Sigma_{\Vt}$ whose one-dimensional cones, denoted by $\Sigma_{\Vt}(1)$, are generated by the non-zero lattice points in $\Delta^\circ$, such that $\Sigma_{\Vt}$ is simplicial and $\Vt$ is a Gorenstein orbifold with terminal singularities.  Such a refinement is known as a maximal projective subdivision.\footnote{There may be many possible maximal projective subdivisions, each corresponding to a different phase of the GLSM (i.e. a different cone in the secondary fan).}  
It may be shown that the singular locus of a Gorenstein orbifold with terminal singularities has codimension $\ge 4$.\footnote{This is nicely discussed in appendix A.2 of~\cite{Cox:2000vi}.} When $d=4$, the singular locus consists of points, and a generic hypersurface $M\subset \Vt$ is smooth.

We will have use for two types of projective subdivisions: a maximal
projective subdivision $\Sigma_{V^{\text{max}}}$, where the one-dimensional cones are generated by
all non-zero lattice points of $\Delta^\circ$, and a minimal
projective subdivision $\Sigma_{V^{\text{min}}}$, where $\Sigma_{V^{\text{min}}}(1)$ is
generated by the non-zero lattice points in $\Delta^\circ$ that are
not contained in the relative interior of any facet $\varphi^\circ
\subset \Delta^\circ$.  The former is a subdivision of the
latter, so that $V^{\text{max}}$ is a resolution of singularities of
$V^{\text{min}}$.  The two only differ in the singular points
corresponding to the non-empty facets of $\Delta^\circ$.  These singular points
in $V^{\text{min}}$ are disjoint from a generic hypersurface $M
\subset V^{\text{min}}$, as are the corresponding exceptional divisors
in $V^{\text{max}}$ from a generic Calabi-Yau hypersurface $\Mt \subset
V^{\text{max}}$.  Thus, when we use these different subdivisions to
construct GLSMs for a Calabi-Yau manifold $M$, we should find two
different descriptions of the same IR physics.  We will refer to the
corresponding GLSMs as ``maximal'' and ``minimal'' models.

\subsection{The Audin/Cox homogeneous coordinate ring}
A $d$-dimensional complete simplicial toric variety $V$ with fan $\Sigma_V \subset N_{\R}$ can be presented as a holomorphic quotient,
\begin{equation}
V \simeq \frac{\C^n - F}{G},
\end{equation}
where $F$ is the ``exceptional set'', $G \simeq (\C^\ast)^{n-d} \times H$ for some finite abelian group $H$, 
and the quotient is the usual geometric quotient~\cite{Cox:1992bob}.

To describe the construction, denote the generators of the
one-dimensional cones by $\rho\in \Sigma_V(1)$, and set $n =
|\Sigma_V(1)|.$ To each $\rho$ we associate a coordinate $Z_\rho$ on
$\C^n$ and a generator of the polynomial ring $S =
\C[Z_{\rho_1},\ldots,Z_{\rho_n}]$.  The Cox ideal $B({\Sigma_V})$ is
defined by
\begin{equation}
B({\Sigma_V}) = \la \prod_{\rho \not\subset \sigma} Z_\rho ~|~\sigma \in \Sigma_V \ra \subset S,
\end{equation}
and the exceptional set $F$ is the subvariety defined by the vanishing of $B({\Sigma_V})$.  When $V$ is simplicial 
$F$ is given by a union of intersections of coordinate hyperplanes in $\C^n$, where for each collection $\{\rho_i\}_{i\in I}$
that does not belong to a full-dimensional cone in $\Sigma_V$, $F$ includes
$\cap_{\rho_i\in I} \{Z_{\rho_i} = 0\}$ as a component.


The group $G$ is determined by the exact sequence
\begin{equation}
\xymatrix{1 \ar[r] & G \ar[r] & (\C^\ast)^{n} \ar[r]^-{\rhot} &T_N \ar[r] & 1,} 
\end{equation}
where $T_N \simeq \C^\ast \otimes N \simeq (\C^\ast)^d$ is the algebraic torus contained in $V$, and
the map $\rhot$ is given by
\begin{equation}
\rhot :  (t_1, \cdots, t_n) \mapsto  (\prod_\rho t_\rho^{\rho^1}, \prod_\rho t_\rho^{\rho^2},\cdots, \prod_\rho t_\rho^{\rho^d} ).
\end{equation}
In other words, $G$ is ``everything in the big torus that does not survive to the little torus.''
The natural action of the big torus 
on $\C^{n}$, 
\begin{equation}
(t_{\rho_1},t_{\rho_2},\ldots,t_{\rho_n})\cdot(Z_{\rho_1},Z_{\rho_2},\ldots,Z_{\rho_n}) = (t_{\rho_1} Z_{\rho_1}, t_{\rho_2} Z_{\rho_2}, \ldots, t_{\rho_n} Z_{\rho_n}),
\end{equation}
induces an action of $G$ 
on the homogeneous coordinate ring $S$, which is thus graded by representations of $G$.  More generally, two Laurent monomials $\prod_\rho Z_\rho^{a_{\rho}}$, $\prod_\rho Z_\rho^{b_{\rho}}$  have equal grading if and only if there exists $m \in M$ such that $a_\rho = b_\rho + \la m, \rho \ra$.   Note that when $V$ is compact, a monomial $\mu \in S$ has degree zero if and only if $\mu \in \C$; otherwise, under the quotient by $G$ $\mu$ would descend to a non-constant holomorphic function on the compact variety $V$.

The homogeneous coordinates lead to a convenient presentation of the
polynomial $P$ that defines the generic Calabi-Yau hypersurface $M \subset V$:
\begin{equation}
\label{eq:P}
P(Z) = \sum_{m\in \Delta\cap M} \alpha_m \prod_\rho Z_\rho^{\la m, \rho \ra + 1}.
\end{equation}
Note that $P(Z)$ is $G$-equivariant.  That is, if an element $g \in G$ acts via 
\begin{equation}
g :\prod_\rho Z_\rho \mapsto \tau \prod_\rho Z_\rho,
\end{equation}
where
$\tau \in \C^\ast$, then $P(g\cdot Z) = \tau P(Z)$.  In particular, $P = 0$ is a well-defined hypersurface in $V$.
In the GLSM it is convenient to choose an explicit basis of integral charges for the $(\C^\ast)^{(n-d)}\subset G$ action.  We will denote this 
basis by $Q^a_\rho$.  That is, the $(\C^\ast)^{(n-d)}$ action will take the form
\begin{equation}
Z_\rho \mapsto \prod_{a=1}^{n-d} t_a^{Q_\rho^a} Z_\rho,~~~t_a \in \C^\ast.
\end{equation}
Under this action the polynomial $P$ has charges $\sum_\rho Q^a_\rho$.

The holomorphic quotient construction leads to an elegant construction of the group of Weil divisors on $V$, denoted by $A_{d-1}(V)$.  $A_{d-1}(V)$ is generated by the torus-invariant divisors $D_\rho$ of the toric variety, modulo linear equivalence encoded by the exact sequence
\begin{equation}
\xymatrix{0 \ar[r] & M \ar[r]^-{\rho} & \Z^{n} \ar[r]^-{f} &A_{d-1}(V) \ar[r] & 0.} 
\end{equation}
The map $\rho$ is
\begin{equation}
\rho : m  \mapsto ( \la m, \rho^1 \ra, \cdots, \la m, \rho^n \ra),
\end{equation}
while the second map sends a point in $\Z^{n}$ to the corresponding divisor:
\begin{equation}
f  : a  \mapsto  \sum_\rho a^\rho D_\rho.
\end{equation}
When $V$ is simplicial, the torus-invariant divisors $D_\rho$ are given by projections of the hyperplanes $\{Z_\rho = 0\}$ under
the quotient. In this case $H^{1,1}(V)$ is generated by the classes $\xi_\rho$ dual to $D_\rho$, and linear equivalence implies $h^{1,1}(V) = n-d$.  The $\xi_\rho$ may be expanded in a basis as $\xi_\rho = \sum_{a} Q_\rho^a \eta_a$.  Moreover, $H^{k,k}(V)$ are the only non-empty cohomology groups, and $H^{\ast,\ast}(V)$ is isomorphic to $\C[\eta_{1},\ldots,\eta_{n-d}]/ I_{\text{S-R}}$, where the Stanley-Reisner ideal $I_{\text{S-R}}$ is obtained by substituting $\xi_\rho$ for $Z_\rho$ in the Cox ideal above.

\subsection{The Automorphism group of $V$}
The homogeneous coordinate ring description leads to a nice presentation of $\Aut V$---the automorphism group of $V$.  This group fits into an exact sequence
\begin{equation}
\xymatrix{1 \ar[r] & G \ar[r] & \Autt V \ar[r]& \Aut V \ar[r] & 1,} 
\end{equation}
and the connected component of the identity  $\Autt_0 V \subset \Autt V$ is described as follows. For each $\rho$, let $S_\rho$ be the  set of monomials in $S$ that have the same degree as $Z_\rho$:
\begin{equation}
\label{eq:srho}
S_\rho = \{\text{monomial}~ \mu \in S ~| ~ \deg [\mu] = \deg [Z_\rho] \}.
\end{equation}
The group $\Autt_0 V$ is generated by elements of  $S_\rho$:  each $\mu \in S_\rho$ yields a one-parameter action on the homogeneous coordinates given by $\delta Z_\rho= \ep \mu$ and $\delta Z_{\rho'\neq \rho} = 0$.  Note that this includes the $(\C^\ast)^n$ action of the ``big torus.''  It follows that the dimension of $\Autt_0 V$ is given by
\begin{equation}
\dim \Autt_0 V = \sum_{\rho} |S_\rho|.
\end{equation}
In the subsequent analysis we will only be interested in $\Aut_0 V$, and for simplicity we will drop the subscript $0$.

When $V$ is constructed from a reflexive polytope as above, there is a nice combinatorial expression for $\dim \Autt V$~\cite{Aspinwall:1993rj}.  To derive this, consider a monomial $\mu \in S_\rho$, $\mu = \prod_\rho Z_\rho^{a_\rho}$.  Suppose $\mu \neq Z_\rho$.   In that case, $a_\rho = 0$, since otherwise $\mu/Z_\rho$ would descend to a non-constant holomorphic function on the compact variety $V$. Recall from above that the degrees of $\mu$ and $Z_\rho$ are 
equal if and only if there exists $m \in M$ such that 
\begin{equation}
a_\rho = \la m, \rho\ra +1 \qquad\text{ and}\qquad a_{\rho'} = \la m,\rho'\ra \qquad \text{for} \qquad \rho'\neq \rho.
\end{equation}
Since $a_\rho = 0$ and $a_{\rho' \neq \rho} \ge 0$ this implies 
\begin{equation}
\la m, \rho \ra = -1,\qquad\text{and}\qquad \la m, \rho' \ra > -1\qquad\text{for}\qquad \rho'\neq \rho.
\end{equation}
If $\rho$ is a vertex of $\Delta^\circ$, these equations imply that $m$ is an interior lattice point of $\varphi_\rho \subset \Delta$, the facet dual to $\rho$. If $\rho$ is not a vertex, then the list of $\rho'$ includes all the vertices, and the inequalities would imply 
$m=0$, in contradiction to the assumption $\mu\neq Z_\rho$.  
Thus, we conclude that
\begin{equation}
\label{eq:dimautt}
\dim \Autt V = n + \sum_{\varphi} \ell^\ast(\varphi) \qquad\text{and}\qquad \dim \Aut V = d + \sum_{\varphi} \ell^\ast(\varphi).
\end{equation}
where the $\varphi$ denote the facets of $\Delta$, and we have used a common notation  $\ell^\ast(P)$ for the number of lattice points in the relative interior of a closed set $P$.  In what follows we will also have use for $\ell(P)$---the number of lattice points in a closed set $P$.

\subsection{Batyrev mirror pairs}
We now have all the ingredients to state the Batyrev mirror correspondence.   Since reflexive polytopes come in pairs, exchanging the roles of $\Delta,\Delta^\circ$ and refining fans as necessary, the construction actually yields a pair of Calabi-Yau hypersurfaces $M,M^\circ$.  It is natural to conjecture that these are mirror, and a check of the proposal is the computation of the Hodge numbers.\footnote{In $d=4$ these are just the usual Hodge numbers of $M$;  when $d>4$ $M$ is generally singular, and one must use the stringy Hodge numbers.}

Let us start with $h^{1,1}(M)$.  A subset of K\"ahler classes on $M$ is obtained by puling back K\"ahler classes on $V$.  If $\Sigma_V$ is a maximal projective subdivision, then $h^{1,1}(V) = \ell(\Delta^\circ) - 1 -d.$  However, not all of the corresponding divisors intersect $M$, since  a divisor corresponding to an interior lattice point of a facet  $\varphi^\circ \subset \Delta^\circ$ is disjoint from $M \subset V$, and its dual class does not contribute to $h^{1,1}(M)$.  Consequently,
\begin{equation}
\hk{M} = \ell(\Delta^\circ) - 1 -d - \sum_{\varphi^\circ} \ell^\ast(\varphi^\circ).
\end{equation}
The same number would have been obtained had we used a minimal projective subdivision.  This is not the whole story, as some of the divisors on $V$ become reducible when restricted to $M$ and  lead to additional generators of $H^{1,1}(M)$.  The full Hodge number is given by
\begin{equation}
h^{1,1} (M) = \hk{M} + \sum_{\Theta^\circ} \ell^\ast(\Theta^\circ) \ell^\ast(\widehat{\Theta^\circ}),
\end{equation}
where $\Theta^\circ \subset \Delta^\circ$ is a codimension-two face of $\Delta^\circ$, and $\widehat{\Theta^\circ} \subset \Delta$ is its dual.
While the correction term does not have the intuitive interpretation of $\hk{M}$, it is still determined by the combinatorics of $\Delta,\Delta^\circ$.

Next, consider the deformations of complex structure, which are counted by $h^{d-2,1}(M)$.  Some of these may be obtained  by varying the coefficients in the defining polynomial.  Naively, this would lead to $\ell(\Delta)$ parameters, but rescaling $P$ by a constant does not change the hypersurface, nor do changes in the parameters of $P$ that may be undone by an action of $\Aut(V)$. Using eqn.~(\ref{eq:dimautt}), we have
\begin{equation}
\hc{M} = \ell(\Delta) - 1 - d -\sum_{\varphi} \ell^\ast(\varphi).
\end{equation}
Again, this is not the whole story, since in general $M$ will have non-polynomial deformations of complex structure.  The correction to $h^{d-2,1}(V)$ has been computed, leading to
\begin{equation}
h^{d-2,1} (M) = \hc{M} +  \sum_{\Theta} \ell^\ast(\Theta) \ell^\ast(\widehat{\Theta}).
\end{equation}

The Hodge numbers for $M^\circ$ are obtained by exchanging the roles of $\Delta$ and $\Delta^\circ$, with the expected result
$h^{1,1}(M) = h^{d-2,1}(M^\circ)$ and $h^{d-2,1}(M) = h^{1,1}(M^\circ)$.  However, note that there is a refinement of this statement:
\begin{equation}
\hc{M} = \hk{M^\circ}, \qquad \hc{M^\circ} = \hk{M}.
\end{equation}
This observation is the basis for the monomial-divisor mirror map~\cite{Aspinwall:1993rj} and an inspiration for our attempt to find a mirror map for the (0,2) GLSM parameters.

\section{The Lagrangian and Its (0,2) Deformations} \label{s:action}
In this section we review the structure of the GLSM Lagrangian for the M-Model~\cite{Witten:1993yc}.  We begin with (2,2) supersymmetry.  The field content includes $1+n$ $(2,2)$ chiral superfields $\Phi_0, \Phi_{\rho}$, where in the maximal case $\rho$ runs over the non-zero lattice points in $\Delta^\circ$, while in the minimal case $\rho$ is further restricted not to lie in the interior of a facet. These matter fields are minimally coupled to $r=n-d$ gauge fields $V_a$ with twisted chiral superfield-strengths $\Sigma_a$.  In addition to the usual flat D-terms, the action is determined by the chiral and twisted chiral superpotentials:
\begin{eqnarray}
{\cL}_W &=& \int~d\theta^+ d\theta^- \Phi_0 P(\Phi) + \text{h.c.}, \nonumber\\
{\cL}_{\Wt} & = & -\frac{1}{4\pi\sqrt{2}} \int d\theta^+ d\thetab^- \log (q_a) \Sigma_a + \text{h.c.}. 
\end{eqnarray}
Here the $q_a$, $ a = 1,\ldots, r$ are the algebraic coordinates on the complexified K\"ahler moduli space for the toric variety V; $P(\Phi)$ is given by the combinatorial data as above; and $\Phi_0$ has charges $Q^a_0 = -\sum_\rho Q^a_\rho$, as required by gauge invariance.  Evidently, the minimal theory has $r$ obvious complexified K\"ahler deformations, which is exactly $\hk{M}$.  The geometric discussion above suggests that in the maximal theory, $\sum_{\varphi^\circ} \ell^\ast(\Gamma^\circ)$ of the $r$ complexified K\"ahler parameters are redundant.  We will discuss this point in greater detail below.

The coefficients of $P(\Phi)$ in ${\cL}_W$ are holomorphic parameters of the GLSM.  Not all of these lead to parameters in the IR theory, since holomorphic field redefinitions can shift some of them into the (presumed irrelevant) D-terms.  The redefinitions compatible with (2,2) supersymmetry and gauge invariance are given by
\begin{equation}
\delta \Phi_0 = \ep_0 \Phi_0,~~~ \delta \Phi_\rho = \ep_{\mu} \mu,~~~ \mu \in S_\rho.
\end{equation}
Thus, the number of field redefinitions is $1 + \dim \Autt V$.  However, gauge invariance of ${\cL}_{W}$ implies that $n-d$ of these redefinitions simply leave the superpotentials invariant.  Thus, the number of complex structure deformations described by the GLSM is 
\begin{equation}
N^{\text{c-x}} = \#(P)  -1 -\dim\Aut V = \ell(\Delta) - 1 -d -\sum_\Gamma \ell^\ast(\Gamma) = \hc{M}.
\end{equation}
Note that $N^{\text{c-x}}$ is the same whether computed in the maximal or minimal GLSM.

The holomorphic terms in the GLSM Lagrangian are particularly convenient for describing the toric and polynomial deformations of the M-model.  This leads to the suspicion that in the more general (0,2) setting the GLSM will also select a natural set of deformations.  To see if this is so, we must first describe the natural set of (0,2) deformations of the (2,2) models.

\subsection{(0,2) Superfields}
Following~\cite{Witten:1993yc}, our first step is to recast the (2,2) theory in terms of (0,2) superfields.  The results are summarized (in Wess-Zumino gauge) by
\begin{eqnarray}
V_a^{(2,2)} & \mapsto & V_{a,-} ~;~ v_{a,+}, \nonumber\\
\Sigma_a^{(2,2)} & \mapsto & \Sigma_a ~;~ \Upsilon_a, \nonumber\\
\Phi_\rho^{(2,2)} &\mapsto & Z_\rho ~;~  \Gamma^\rho, \nonumber\\
\Phi_0^{(2,2)} &\mapsto & Z_0 ~;~  \Gamma^0,
\end{eqnarray}
where $\Sigma_a, Z_\rho$ are bosonic chiral multiplets, $\Upsilon_a$ are chiral Fermi gauge field-strength multiplets, and the $\Gamma^\rho$ are Fermi multiplets obeying
\begin{equation}
\cDb_+ \Gamma^\rho = i \sqrt{2} \sum_{a=1}^{r} \Sigma_a Q^a_\rho Z_\rho.
\end{equation}
Here $\cDb_+$ is a (0,2) superspace derivative.  The (2,2) superpotentials combine into a single (0,2) superpotential $\cL_{\cW}$ that encodes the holomorphic terms in the action:
\begin{equation}
\cL_{\cW} = \int d \theta^+~\left\{ \frac{\log(q_a)}{8\pi i} \Upsilon_a + \Gamma^0 P(Z) + Z_0 \sum_\rho \Gamma^\rho P_{,\rho} \right\} + \text{h.c.},
\end{equation}
where $P_{,\rho} = \p P/\p Z_\rho$.
Since the $\Gamma^\rho$ are not chiral, this action is not obviously supersymmetric.  The condition for supersymmetry (i.e. for $\cL_{\cW}$ to be chiral) is 
\begin{equation}
Z_0\sum_{a=1}^{r}  \Sigma_a \left[ Q_0^a P(Z) + \sum_\rho Q_\rho^a Z_\rho P_{,\rho} \right] = 0.
\end{equation}
But $Q_0^a = -\sum_i Q_i^a$, and gauge invariance implies that $P$ is quasi-homogeneous, so that the last term just yields $+\sum_\rho Q_\rho^a P(\Phi)$. The terms in the bracket cancel for all $a$, and the theory is supersymmetric after all.

In addition to the gauge symmetry, the action has two important global $\GU(1)$ symmetries: the $\GUR$ gives charge $+1$ to $\theta^+$ and should become the $\GU(1)$ current in the right-moving $N=2$ algebra; the $\GUL$ is a global symmetry from the point of view of (0,2) superspace, but will become the $\GU(1)$ current in the left-moving $N=2$ algebra.  The charges of the superfields, determined up to gauge transformations, are given in table~\ref{table:charges}.  Any (0,2) deformations should preserve these symmetries.
\begin{table}[t]
\begin{center}
\begin{tabular}{|c|c|c|c|c|c|c|c|}
\hline
$~$			&$\theta^+$	&$Z_\rho$	&$\Gamma^\rho$	&$Z_0$		&$\Gamma^0$	&$\Sigma_a$	&$\Upsilon_a$\\ \hline
$\GUR$		&$1$			&$0$		&$0$				&$1$			&$1$			&$1$			&$1$\\ \hline
$\GUL$		&$0$			&$0$		&$-1$			&$1$			&$0$			&$-1$		&$0$\\ \hline
\end{tabular}
\end{center}
\caption{Symmetry Charges}
\label{table:charges}
\end{table}

\subsection{(0,2) Parameters}
The (2,2) action above has an obvious (0,2) generalization: write a general chirality constraint for the $\Gamma^\rho$ and use the most general superpotential $\cL_{\cW}$ consistent with gauge invariance, $\GUR\times\GUL$ symmetries, and supersymmetry.  Restricting to terms polynomial in the fields, the solution is
\begin{equation}
\cDb_+ \Gamma^0 = \sum_a \Sigma_a E^{a0} Z_0,~~~\cDb_+ \Gamma^\rho = E^\rho(\Sigma,Z) = \sum_a \Sigma_a E^{a\rho}(Z),
\end{equation}
and
\begin{equation}
\cL_{\cW} = \int d \theta^+~\left\{ \frac{\log(q_a)}{8\pi i} \Upsilon_a + \Gamma^0 P(Z) + Z_0 \sum_\rho \Gamma^\rho J_\rho(Z) \right\} + \text{h.c.},
\end{equation}
where $E^{a0}$ is a set of constants, the polynomials $E^{a\rho}$ have gauge charges $Q^a_\rho$, and the $J_\rho$ have charges $-Q^a_0-Q^a_\rho$.  The superpotential is chiral if and only if
\begin{equation}
\label{eq:susyconstr}
E^{a0} P(Z) + \sum_\rho E^{a\rho}(Z) J_\rho(Z) = 0~~~\text{for all}~~~a.
\end{equation}
We have assumed that $Z_0$ is the unique monomial in the $Z_0,Z_\rho$ with charges $Q_0^a$.  This is reasonable for a hypersurface but will be modified in the complete intersection case.

Working in a large radius phase of the GLSM, it is not hard to see that this indeed corresponds to deformations of the tangent bundle~\cite{Witten:1993yc,Distler:1995mi} described by the complex
\begin{equation}
\xymatrix{ 0\ar[r] &\cO^{r}|_M \ar[r]^-{E^{a\rho}} &\oplus_\rho\cO(D_\rho)|_{M}  \ar[r]^-{J_\rho} & \cO(\sum_\rho D_\rho)|_{M}\ar[r] & 0 }.
\end{equation}
The supersymmetry condition of eqn.~(\ref{eq:susyconstr}) ensures this is indeed a complex, and the quotient
$\cF = \ker J / \im E$ defines a rank $3$ holomorphic bundle over $M$.  Setting $E^{a\rho}$ and $J_\rho$ to their (2,2) values determines the constants $E^{a0}$ to be the charges, and $\cF = T_M$.

\subsection{Field Redefinitions} \label{ss:redefs}
Just as in the (2,2) case, we do not expect that all of the parameters in the $E$ and $J$ correspond to deformations, since field redefinitions can be used to absorb some of the parameters into deformations of the presumably irrelevant D-terms.  The field redefinitions should be consistent with supersymmetry, global symmetries and gauge invariance.  If we also assume, as in the (2,2) case, that the redefinitions are polynomial in the fields, the possibilities are restricted to
\begin{equation}
\begin{array}{lcl}
\delta Z_0 = u Z_0, &\quad&  \delta \Gamma^0 = v \Gamma^0, \\
\delta Z_\rho = \sum_{\mu \in S_\rho} \ep_\mu \mu,   &\quad& \delta \Gamma^\rho = \sum_{\rho'} \sum_{\nu \in S_{\rho\rho'}} \eta_\nu \nu \Gamma^{\rho'}, \\
\delta \Sigma_a = G_{a}^{b} \Sigma_b, &\quad& \delta \Upsilon_a = 0.
\end{array}
\end{equation}
Here $u,v,\ep_\mu,\eta_\nu$ are complex parameters; $G^a_b$ is a field-independent $\GL(r,\C)$ matrix;
$S_\rho$ is as in eqn.~(\ref{eq:srho}); and
\begin{equation}
S_{\rho\rho'} = \{\text{monomial}~ \nu \in S~|~ \deg[\nu] = \deg[Z_\rho] - \deg[Z_{\rho'}]\}.
\end{equation}
Since these redefinitions are holomorphic, the $1+r$ redefinitions corresponding to $\GUL$ symmetry and gauge transformations will leave $\cL_{\cW}$ and $E^{a\rho}$ invariant.
When these redefinitions were first considered in~\cite{McOrist:2008ji}, a mistake was made in the form of $\delta \Gamma^\rho$.  Although this error does not affect models where the monomials in $S_\rho$ are linear in the fields, it does lead to extra redefinitions in more general models.

As long as the $\mu$ and $\nu$ monomials are linear in the $Z_\rho$, these redefinitions have well-defined Jacobians, at least at the level of the bosonic zero modes; however, it is not so obvious that this holds for non-linear $\mu$ and $\nu$.  To see that the non-linear terms cause no trouble, consider the redefinition of the $\Gamma^\rho$:  
\begin{equation}
\Gammat^\rho = \Gamma^\rho+\delta\Gamma^\rho= \sum_{\rho'} U^\rho_{\rho'}(Z) \Gamma^{\rho'}.
\end{equation}
At first sight, since $U$ depends on $Z$, no matter how small the parameters $\eta_\nu$ in $\delta\Gamma$ are taken to be, $\det U$ might vanish for some non-zero $Z$.  However, since $\det U$ is a polynomial in the $Z$ of gauge charge zero, it must actually be independent of $Z$, since otherwise it would descend to a non-constant function on the compact variety $V$.  Since $\det U$ is $Z$-independent, sufficiently small $\eta_\nu$ parameters will ensure $\det U \neq 0$.  A similar argument may be given for the redefinitions of the $Z_\rho$.

This is a good point to note that in general the K\"ahler parameters will not be invariant under generic (0,2) redefinitions of the matter fields.  This is easily seen in the geometric phase in an instanton expansion, where the field-independent determinant factors from redefinitions of the fermion measure will shift the action by terms proportional to the instanton number.  This fact is sure to play an important role in any definition of a (0,2) GLSM mirror map.

\section{Combinatorics of Deformations} \label{s:combin}
Having described the $E$ and $J$ parameters and the redefinitions, we will now give combinatorial expressions for the holomorphic (0,2) deformations of the GLSM.

\subsection{E and J parameters}
The polynomial $P(Z)$ defining the hypersurface contains $\ell(\Delta)$ para\-meters---the coefficients $\alpha_m$ of the monomials $\prod_\rho Z_\rho^{\la m,\rho \ra +1}$.  On the (2,2) locus the $J_\rho$ are given by 
\begin{equation}
J_{\rho} = P_{,\rho} = \sum_{m \in \Delta \cap M} (\la m, \rho \ra + 1) \alpha_m Z_\rho^{\la m,\rho \ra} \prod_{\rho'\neq\rho} Z_{\rho'}^{\la m, \rho'\ra+1}~~.
\end{equation}
More generally, $J_\rho$ must contain the same set of monomials as $P_{,\rho}$, but their coefficients need not be related to the $\alpha_m$.  Each $\rho \in \Delta^\circ \cap N$ belongs to the relative interior of a face of minimal dimension $\tau_\rho^\circ$.  It is not hard to show that $m \in \Delta$ satisfies $\la m, \rho \ra = -1$ if and only if $m$ belongs to the dual face of $\tau_\rho^\circ$, $\tau_\rho \subset \Delta$.  Thus, the general $J_\rho$ takes the form\footnote{There is a slight abuse of notation here, since $m$ is a {\em lattice} point in $\Delta\backslash\tau_\rho$.  Similarly, $\rho$ is restricted to be a generator of $\Sigma_V(1)$.  We will make use of this abuse when it is not likely to lead to confusion.}
\begin{equation}
J_\rho =  \sum_{m \in \Delta\backslash\tau_\rho} \left[ (\la m,\rho \ra+1)\alpha_m + \beta_m^\rho \right] Z_\rho^{\la m,\rho\ra} \prod_{\rho' \neq \rho} Z_{\rho'}^{\la m,\rho'\ra +1 },
\end{equation}
where the $\beta^\rho_m$ parametrize the difference from the (2,2) locus values.  The number of these parameters is
\begin{equation}
\label{eq:Jnumgen}
\# J = \sum_\rho \sum_{m\in\Delta}  \left[1 - \delta (\la m, \rho \ra + 1) \right] = n \ell(\Delta) - \sum_\rho \ell(\tau_\rho).
\end{equation}

Next, we consider the parameters in  $E^{a0}$ and $E^{a\rho}$. $E^{a0} = (Q^a_0 + \delta^a)$, leading to $r$ parameters $\delta^a$.  To describe the $E^{a\rho}$, it is useful to consider separately the vertices of $\Delta^\circ$.  Let $\Ve$ be the set of vertices of $\Delta^\circ$.  If $\rho \in \Ve$ then $\tau_\rho = \varphi_\rho$, the facet of $\Delta$ dual to $\rho$.  Denoting the relative interior of a set $K \subset M_{\R}$ by $\ri K$, the $S_\rho$ take the form
\begin{equation}
\label{eq:smartsrho}
S_\rho = \begin{cases}
 \{Z_\rho\} \cup \left\{\prod_{\rho'\neq\rho} Z_{\rho'}^{\la m, \rho'\ra} ~|~ m \in \ri\varphi_\rho \right\}, & \rho \in \Ve \\
 \{ Z_\rho \} & \text{otherwise.}  \\
\end{cases}
\end{equation}
Using this decomposition, we can write down the $E^{a\rho}$:
\begin{equation}
E^{a\rho} = \begin{cases}  
(Q^a_\rho+\gamma^{a0}_\rho) Z_\rho + \sum_{m \in\ri\varphi_\rho} \gamma_\rho^{am} \prod_{\rho'\neq\rho} Z_{\rho'}^{\la m,\rho' \ra }, & \rho \in\Ve \\
(Q^a_\rho+\gamma^{a0}_\rho) Z_\rho & \text{otherwise.}
\end{cases}
\end{equation}
Evidently, the total number of the $E$ parameters, $\delta^a$, $\gamma^{a m}_\rho$ and $\gamma^{a0}_\rho$ is
\begin{equation}
\# E  = r\left(1+n+ \sum_{\varphi} \ell^\ast(\varphi) \right)= r (1+\dim\Autt V ).
\end{equation}

Finally, we need to examine the supersymmetry constraint.  Plugging $P$, $E$ and $J$ into eqn.~(\ref{eq:susyconstr}), we find the condition
\begin{equation}
\sum_{m\in\Delta} F^a_m \prod_{\rho} Z_\rho^{\la m,\rho\ra+1}=0,
\end{equation}
 where
\begin{eqnarray}
\label{eq:constsusy}
F^a_m & = & \delta^a \alpha_m + \sum_\rho \sum_{l\in \Delta\backslash\tau_\rho} \left[\gamma^{a0}_\rho (\la l,\rho \ra +1) \alpha_l + Q^a_\rho \beta^\rho_l +\gamma^{a0}_\rho \beta^\rho_l \right] \delta(l-m) \nonumber\\
~ & ~ & + \sum_{\rho\in\Ve} \sum_{l \in \ri\varphi_\rho} \sum_{k \in \Delta\backslash\tau_\rho} \left[\gamma^{a l}_\rho (\la k,\rho\ra +1)\alpha_k+ \gamma^{a l}_\rho \beta^\rho_k\right]\delta(l+k-m).
\end{eqnarray}
This yields $r\ell(\Delta)$ constraints on the parameters, and if the equations are generic enough, they should eliminate $r\ell(\Delta)$ of the parameters.

\subsection{The Redefinitions}
Having described the parameters and the constraint on them in terms of the combinatorics, we will now do the same for the field redefinitions.
The rescalings of $Z_0$ and $\Gamma^0$ are simple enough, as are the $\GL(r,\C)$ rotations of the $\Sigma_a$.  Next, we turn to the redefinitions of the $Z_\rho$. Using the decomposition of the $\rho$ into vertices and other points leads to
\begin{equation}
\label{eq:zredef}
Z'_\rho = \begin{cases}
   (1+\ep^0_\rho) Z_\rho + \sum_{m \in \ri\varphi_\rho}  \ep^m_\rho \prod_{\rho'\neq\rho} Z_{\rho'}^{\la m,\rho'\ra}, & \rho \in \Ve \\
   (1+\ep^0_\rho) Z_\rho & \text{otherwise}.\\
\end{cases}
\end{equation}
The number of parameters contained in these is 
\begin{equation}
\#\delta Z = n+\sum_{\varphi} \ell^\ast(\varphi) = \sum_\rho |S_\rho| =  \dim\Autt V.
\end{equation}
Note that the non-linear redefinitions only affect the $Z_\rho$ corresponding to the vertices of $\Delta^\circ$.

Next, we consider the redefinitions of the $\Gamma^\rho$.  We again consider $\rho\not\in\Ve$ and $\rho \in \Ve$ separately and find
\begin{equation}
{\Gamma'}^\rho = (1+\eta^\rho_0) \Gamma^\rho \qquad\text{for}\qquad \rho \not\in \Ve,
\end{equation}
and when $\rho \in \Ve$
\begin{equation}
{\Gamma'}^\rho =(1+\eta^\rho_0) \Gamma^\rho+\sum_{m\in\ri\varphi_\rho} \sum_{\rho'\neq \rho} \left[1-\delta(\la m,\rho'\ra)\right] \eta^{m\rho}_{\rho'} Z_{\rho'}^{-1} \prod_{\rho''\neq\rho} Z_{\rho''}^{\la m, \rho'' \ra }\Gamma^{\rho'}.
\end{equation}
These redefinitions depend on 
\begin{equation}
\#\delta\Gamma = n + \sum_{\rho\in\Ve} \sum_{m \in \ri\varphi_\rho} \sum_{\rho'\neq\rho} \left[1 - \delta(\la m,\rho'\ra)\right]
\end{equation}
parameters.  This can be recast as
\begin{equation}
\#\delta\Gamma = \#\delta Z + \sum_{\varphi}\sum_{m\in\ri\varphi}\left[n-2- \sum_{\rho} \delta(\la m,\rho\ra)\right],
\end{equation}
showing that in general  $\#\delta\Gamma \ge \#\delta Z$.  Note that when $\Delta^\circ$ describes
a product of projective spaces, $\#\delta\Gamma = \#\delta Z$.  This is easy to see by thinking about the monomials,
and may also be shown directly from the formula for $\#\delta\Gamma$.

\subsection{The Number of Deformations}
We are now ready to count the deformations. In addition to the parameters described above, we should also add in the $r$ complexified K\"ahler deformations.  Although naively there are $r$ such parameters, for each $\rho$ contained in a relative interior of a facet of $\Delta^\circ$, one of the K\"ahler parameters should be redundant.  Let us call the number of these redundant K\"ahler parameters $w$.  Adding these to the numbers described above, we have a count of the deformations of the M-model GLSM:
\begin{eqnarray}
N(M) & = & r -w +\#P ~+ ~\underbrace{r(1+\#\delta Z)}_{\text{E}} +\#J \nonumber\\
~ &~&- \underbrace{r\#P}_{\text{SUSY}} - \underbrace{(2 + r^2 +\#\delta Z+\#\delta\Gamma -1 -r)}_{\text{redefinitions/gauge \&$\GUL$}}\nonumber\\
~ & = & 1-w -(r-1)( r+\#P-\#\delta Z-2) + \#J - \#\delta\Gamma.
\end{eqnarray}
We will distinguish several different choices for the M-model:  the minimal choice, where $w = 0$; the maximal choice, where the $\rho$ include every non-zero lattice point of $\Delta^\circ$; and the special situation where neither $\Delta$ or $\Delta^\circ$ has a facet with an interior lattice point.  To discuss these different choices it is convenient to define
\begin{equation}
\begin{array}{lcl}
X = \ell(\Delta)-1,&\quad & X^\circ = \ell(\Delta^\circ) -1, \\
W = \sum_\varphi \ell^\ast(\varphi), &\quad & W^\circ = \sum_{\varphi^\circ} \ell^\ast(\varphi^\circ).
\end{array}
\end{equation} 
In the case of the minimal M-model, where $w=0$, the general expression reduces to
\begin{eqnarray}
N(M_{\text{min}}) &=& (d+1)(X+X^\circ-W-W^\circ) -d(d+2) \nonumber\\
~ & ~&  -\sum_{\rho}\ell(\tau_\rho) +W+\sum_{\varphi}\sum_{m\in\ri\varphi} \sum_{\rho} \delta(\la m,\rho\ra).
\end{eqnarray}
While the first line is invariant under the exchange of $\Delta,\Delta^\circ$, the second one is
not invariant in any obvious way.  Indeed, computations in examples will show that the second line fails to be mirror symmetric.
For the maximal M-model, where $w=W^\circ$, we find
\begin{eqnarray}
N(M_{\text{max}}) & = & (d+1) (X+X^\circ) -W-W^\circ -d(d+2)  -\sum_{\rho}\ell(\tau_\rho) \nonumber \\
~ &~ & -(d-1) W +\sum_{\varphi}\sum_{m\in\ri\varphi} \sum_{\rho} \delta(\la m,\rho\ra).
\end{eqnarray}
In this case, the first line is mirror symmetric, since 
\begin{equation}
\sum_\rho \ell(\tau_\rho) = \sum_{\rho \in \Delta^\circ\backslash\{0\}} \sum_{m\in\Delta} \delta(\la m, \rho \ra+1)
= \sum_{m \in \Delta\backslash\{0\}} \sum_{\rho\in \Delta^\circ} \delta(\la m, \rho \ra+1) 
= \sum_{m} \ell(\tau^\circ_m).
\end{equation}
However, the second line is again not obviously mirror symmetric, and the difference between the model and its mirror is
\begin{eqnarray}
N(M_{\text{max}})-N(M_{\text{max}}^\circ) &=& (d-1) (W^\circ-W) + \sum_{\varphi} \sum_{m\in\ri\varphi} \sum_{\rho\not\in\varphi^\circ} \delta(\la m,\rho \ra) \nonumber\\
~&~&-\sum_{\varphi^\circ} \sum_{\rho\in\ri\varphi^\circ} \sum_{m\not\in\varphi} \delta(\la m,\rho \ra).
\end{eqnarray}
 Comparing $N(M_{\text{max}})$ and $N(M_{\text{min}})$, we note that the two are in general not the same:
\begin{equation}
N(M_{\text{max}}) - N(M_{\text{min}}) = (d-1) W^\circ + \sum_{\varphi} \sum_{m \in \ri \varphi} \sum_{\varphi^\circ} \sum_{\rho \in \ri \varphi^\circ}\delta(\la m,\rho \ra ) \ge 0.
\end{equation}
Contrary to the counting of (2,2)-preserving deformations, it appears that in general the maximal GLSM yields more (0,2) deformations than the minimal one.

\subsection{Reflexively Plain Pairs}
It is clear that $N(M_{\text{max}}) = N(M_{\text{min}})$ when $W^\circ = 0$.  Since in this case the polytope $\Delta^\circ$ has no interior lattice points in any facet, we will say the polytope $\Delta^\circ$ is plain.  When $W =0$ as well, so that $\Delta$ is also a plain polytope, we will call the pair $(\Delta^\circ,\Delta)$ a reflexively plain pair.  In this case the counting of deformations  is mirror symmetric: 
\begin{equation}
N(M) = (d+1) (X+X^\circ) -d(d+2) - \sum_\rho\ell(\tau_\rho) = N(M^\circ).
\end{equation}
We can split this into the (2,2) and (0,2) parameters:
\begin{eqnarray}
N^{(2,2)}(M) &=& (X-d) + (X^\circ -d),\nonumber\\
N^{(0,2)}(M) &=& d(X+X^\circ-d) -\sum_\rho \ell(\tau_\rho).
\end{eqnarray}
These theories are candidates for (0,2) GLSM mirror pairs.  The reader may, as did the authors, wonder how many such models exist.  In four dimensions there is a classification of reflexive polytopes~\cite{Kreuzer:2000xy,Kreuzer:2002uu}.  A scan through the PALP database of the $473,800,776$ four-dimensional reflexive polytopes identified $6,677,743$ reflexively plain pairs and $5518$ self-dual plain polytopes.  Although this amounts to just under $3\%$ of the database, the absolute number is large enough that we can be sure of finding plenty of interesting examples!  We intend to return to a study of the reflexively plain models in the near future; for now, however, we will explore in some more detail the mismatch between the GLSM deformations in more general theories.  We will start by computing the discrepancy in a number of examples.

\subsection{Examples}

A family of models that shows enough of the interesting structure consists of Calabi-Yau hypersurfaces in $\P^{n_1} \times \P^{n_2}\times\cdots\times \P^{n_k}$.  In this case the minimal and maximal GLSMs are identical, $d = \sum_{\alpha} n_\alpha$, and the rank of the gauge group is just $r = k$.
Counting the monomials in P, we find
\begin{equation}
\#(P) = \ell(\Delta) = \prod_{\alpha=1}^k \binom{2n_\alpha +1}{n_\alpha} \equiv X + 1.
\end{equation}
The field redefinitions are also easy to count.  Since the $S_\rho$ are linear in the fields
\begin{equation}
\#\delta Z  = \#\delta \Gamma = \sum_{\alpha=1}^k (n_\alpha+1)^2 = k + d + W,
\end{equation}
where, as above, $W$ is the number of interior facet lattice points in $\Delta$.  It is not hard to see that
$\Delta$ has $\sum_{\alpha} (n_\alpha+1)$ facets, each of which has $n_\alpha$ interior lattice points.

Finally, we need the number of $J$-parameters.  Let $\rho$ correspond to a coordinate of $\P^{n_\alpha}$.  Then the number of monomials in $J_\rho$ is
the number of monomials of multi-degree $(n_1+1,\ldots, n_\alpha,\ldots n_{k}+1)$.   Note that each $\rho$ corresponding to $\P^{n_\alpha}$
has the same number of monomials.  Summing over the $k$ $\P^{n_\alpha}$ factors yields
\begin{equation}
\# (J) =  \sum_{\alpha=1}^k (n_\alpha+1) \binom{2n_\alpha}{n_\alpha} \prod_{\beta\neq\alpha} \binom{2n_\beta+1}{n_\beta} \equiv Z.
\end{equation}
Putting these terms into the general expression, we conclude that 
\begin{equation}
N(M_\text{min}) = N(M_\text{max}) = Z + d (k-2) -(k-1) X + (k-2) W.
\end{equation}
As was noted in~\cite{McOrist:2008ji}, for the case of the quintic in $\P^4$ and bi-cubic in $\P^2\times\P^2$, the number obtained here match $h^{1,1} + h^{2,1}+\dim H^1(\End T_M)$ computed in, for instance,~\cite{Hubsch:1992nu}.

Next, we turn to the minimal mirror.  The fan has $X - W$ one-dimensional cones, and thus the  rank of the gauge group is
\begin{equation}
r^{\circ} = X- d - W.
\end{equation}
Since $\Delta^\circ$ is plain, the automorphism group of the mirror toric variety is minimal, and
\begin{equation}
\#\delta Z = \#\delta \Gamma = X - W.
\end{equation}
Of course, 
\begin{equation}
\#P^\circ = \sum_{\alpha} (n_\alpha+1) + 1 = 1+k + d.
\end{equation}
The counting of $J^\circ$ is easiest to do in two steps. First, note that the $\rho=0$ monomial in $P^\circ$ contributes to every $J^\circ_m$, and therefore yields $X-W$ parameters.  Now suppose we pick some monomial in $P^\circ$ with $\rho\neq 0$.  This contributes to every $J^\circ_m$ with $m$ that is not in the face dual to $\rho$.   This is just the number of monomials in $P$ with $m\neq 0$ and $m\not \in \ri \varphi$ for any facet $\varphi\subset \Delta$, that have at least one power of $Z_\rho$.  If in the M-model $\rho$ corresponds to a coordinate on $\P^{n_\alpha}$, then the $\rho$ monomial in $P^\circ$ appears in
\begin{equation}
\prod_\beta \binom{2 n_\beta+1}{n_\beta} - 1 -\sum_\beta n_\beta(n_\beta+1) - \left[\binom{2 n_\alpha}{n_\alpha-1} \prod_{\beta\neq\alpha} \binom{2 n_\beta+1}{n_\beta} - n_\alpha \right]
\end{equation}
of the $J^\circ_m$. Summing over $\rho\in \Delta^\circ \backslash \{0\}$, we find
\begin{equation}
\#(J^\circ) = Z+X - (d+k) (W+1).
\end{equation}
Putting these ingredients into the general formula, we obtain
\begin{equation}
N(M^\circ_{\text{min}}) =  Z+ d(k-2)  - (k-1) X - dW.
\end{equation}

Finally, we consider the maximal mirror model.   The details are much the same as in the minimal mirror model, and we find
\begin{equation}
\begin{array}{lcl}
w^\circ = W, &\quad & r^\circ =  X - d,\\
\#P^\circ = d + k+1, & \quad&\#\delta Z^\circ = \#\delta\Gamma^\circ = X, \\
 \#J^\circ = Z+X -d -k. &\quad&~
\end{array}
\end{equation}
Putting this together, 
\begin{equation}
N(M^\circ_{\text{max}}) =  Z+d(k-2)  -(k-1) X - W.
\end{equation}
Comparing the answers, 
\begin{eqnarray}
N(M) - N(M^\circ_{\text{max}}) & = & (k-1) W, \nonumber\\
N(M) - N(M^\circ_{\text{min}} ) & = & (k-2+d) W.
\end{eqnarray}
We see that  unless $k=1$ (the case of a single projective space), 
$N(M) \neq N(M^\circ_{\text{max}})$.

Another simple set of examples is constituted by hypersurfaces in $d=4$ weighted projective spaces with appropriately resolved singularities.  Taking the $h^{1,1} = 2$ examples studied in~\cite{Hosono:1993qy}, we find the results given in table~\ref{table:weighted2}.  Results for models with $h^{1,1} = 3$ are given in table~\ref{table:weighted3}.  Several qualitative features can be discerned from these numbers.  First, as expected $N(M_{\text{max}}) \ge N(M_{\text{min}})$.  Second, comparing the differences $N(M_{\text{min}})-N(M_{\text{min}}^\circ)$ and $N(M_{\text{max}})-N(M_{\text{max}}^\circ)$, there is no sign of one being a ``better match'' than the other.  Finally, it is rather apparent that the discrepancies become more dramatic as the number of lattice interior facet points increases.

\begin{table}[h]
\begin{center}
\begin{tabular}{|c|c|c|c|c||c|c|||c||c|}
\hline
Model			&$r$		&$\#P$	&$\#\delta Z$	&$\#\delta\Gamma$	&$\#J$	&$\ell^\ast(\varphi)$		& $N(M)$ 
\\ \hline
$\P^4_{1,1,2,2,2}$	&$2$		&$105$	&$23$		&$35$		&$380$		&$17$				&$264$
\\ \hline
min. mir.			&$83$	&$7$		&$87$		&$87$		&$376$		&$0$					&$208$
\\ \hline
max. mir.			&$100$	&$7$		&$104$		&$104$		&$478$		&$0$					&$259$
\\ \hline\hline
$\P^4_{1,1,2,2,6}$	&$2$		&$171$	&$46$		&$118$		&$664$		&$40$				&$422$
\\ \hline
maximal			&$3$		&$171$	&$47$		&$149$		&$834$		&$40$				&$435$
\\ \hline
min. mir.			&$126$	&$8$		&$131$		&$240$		&$717$		&$1$					&$353$
\\ \hline
max. mir.			&$166$	&$8$		&$171$		&$310$		&$997$		&$1$					&$483$
\\ \hline\hline
$\P^4_{1,2,2,3,4}$	&$2$		&$89$	&$20$		&$33$		&$320$		&$14$				&$219$
\\ \hline
min. mir.			&$70$	&$7$		&$74$		&$74$		&$318$		&$0$					&$176$
\\ \hline
max. mir.			&$84$	&$7$		&$88$		&$88$		&$402$		&$0$					&$218$
\\ \hline\hline
$\P^4_{1,2,2,2,7}$	&$2$		&$141$	&$35$		&$78$		&$512$		&$29$				&$329$
\\ \hline
maximal			&$4$		&$141$	&$37$		&$123$		&$792$		&$29$				&$350$
\\ \hline
min. mir.			&$107$	&$9$		&$113$		&$282$		&$692$		&$2$					&$305$
\\ \hline
max. mir.			&$136$	&$9$		&$142$		&$354$		&$924$		&$2$					&$407$
\\ \hline\hline
$\P^4_{1,1,1,6,9}$	&$2$		&$376$	&$105$		&$331$		&$1533$		&$99$				&$932$
\\ \hline
maximal			&$5$		&$376$	&$108$		&$575$		&$2658$		&$99$				&$997$
\\ \hline
min. mir.			&$272$	&$10$	&$279$		&$1045$		&$2133$		&$3$					&$818$
\\ \hline
max. mir.			&$371$	&$10$	&$378$		&$1385$		&$3024$		&$3$					&$1171$
\\ \hline
\end{tabular}
\caption{Some weighted projective spaces with $h^{1,1} = 2$. Here $\ell^\ast(\varphi)$ is the number of lattice points in the relative interior of facets of $\Delta$.}
\label{table:weighted2}
\end{center}
\end{table}
\begin{table}[h]
\begin{center}
\begin{tabular}{|c|c|c|c|c||c|c|||c||c|}
\hline
Model			&$r$		&$\#P$	&$\#\delta Z$	&$\#\delta\Gamma$	&$\#J$	&$\ell^\ast(\varphi)$		&$N(M)$ 
\\ \hline
$\P^4_{1,1,1,3,6}$	&$2$		&$225$	&$61$		&$170$		&$898$		&$55$				&$565$
\\ \hline
maximal			&$3$		&$225$	&$62$		&$210$		&$1122$		&$55$				&$584$
\\ \hline
min. mir.			&$165$	&$8$		&$170$		&$315$		&$954$		&$1$					&$476$
\\ \hline
max. mir.			&$220$	&$8$		&$225$		&$409$		&$1339$		&$1$					&$657$
\\ \hline\hline
$\P^4_{1,2,3,3,3}$	&$3$		&$81$	&$20$		&$36$		&$354$		&$13$				&$195$
\\ \hline
min. mir.			&$63$	&$8$		&$67$		&$67$		&$336$		&$0$					&$146$
\\ \hline
max. mir.			&$76$	&$8$		&$80$		&$80$		&$427$		&$0$					&$185$
\\ \hline\hline
$\P^4_{1,3,3,3,5}$	&$3$		&$81$	&$20$		&$34$		&$342$		&$13$				&$185$
\\ \hline
maximal			&$4$		&$81$	&$21$		&$42$		&$422$		&$13$				&$194$
\\ \hline
min. mir.			&$63$	&$9$		&$68$		&$106$		&$390$		&$1$					&$161$
\\ \hline
max. mir.			&$76$	&$9$		&$81$		&$126$		&$494$		&$1$					&$206$
\\ \hline\hline
$\P^4_{1,2,3,3,9}$	&$3$		&$130$	&$37$		&$115$		&$618$		&$30$				&$316$
\\ \hline
maximal			&$4$		&$130$	&$38$		&$139$		&$747$		&$30$				&$326$
\\ \hline
min. mir.			&$95$	&$9$		&$100$		&$182$		&$628$		&$1$					&$259$
\\ \hline
max. mir.			&$125$	&$9$		&$130$		&$235$		&$868$		&$1$					&$356$
\\ \hline\hline
$\P^4_{1,1,2,8,12}$	&$3$		&$335$	&$95$		&$380$		&$1692$		&$88$				&$831$
\\ \hline
maximal			&$6$		&$335$	&$98$		&$598$		&$2694$		&$88$				&$889$
\\ \hline
min. mir.			&$242$	&$11$	&$249$		&$930$		&$2138$		&$3$					&$727$
\\ \hline
max. mir.			&$330$	&$11$	&$337$		&$1233$		&$3018$		&$3$					&$1040$
\\ \hline
\end{tabular}
\caption{Some weighted projective spaces with $h^{1,1} = 3$. Note that the first example has a non-toric divisor. Here $\ell^\ast(\varphi)$ is the number of lattice points in the relative interior of facets of  $\Delta$.}
\label{table:weighted3}
\end{center}
\end{table}

As a final example, we consider one of the reflexively plain pairs identified by the search through the database of four-dimensional reflexive polytopes.  Here the plain polytopes have vertices
\begin{equation}
\Delta^\circ    : \begin{pmatrix} 1 & 0 & 2 & 3 & -6 \\ 0 & 1 & 4 & 3 & -8 \\ 0 & 0 & 5 & 0 & -5 \\ 0 & 0 & 0 & 5 & -5 \end{pmatrix},\quad
\Delta	     : \begin{pmatrix} -1 & -1 & 1 & 1 \\ -1 & -1 & 1 & 2 \\  -1 & -1 &2 & 1 \\ -1 & 4 & -3 & -2 \\ 4 & -1 & -1 &-2\end{pmatrix}.
\end{equation}
$\Delta^\circ$ has a total of $26$ lattice points, while $\Delta$ has no additional non-zero lattice points.  By construction neither has facets with interior lattice points.   Performing the count, we find
\begin{equation}
\begin{array}{lcl}
r  = 21, & \quad & r^\circ = 1, \\
\#P = 6, &\quad & \#P^\circ = 26, \\
\#\delta Z = 25, &\quad & \#\delta Z^\circ = 5, \\
\#\delta\Gamma = 25, & \quad &\#\delta\Gamma^\circ = 5, \\
\#J = 90, &\quad & \#J^\circ = 70,  \\
N(M) = 66, &\quad& N(M^\circ) = 66.
\end{array}
\end{equation}

\section{Discussion} \label{s:discussion}
The simplest interpretation of our results is that (0,2) deformations expressible as parameters in the (0,2) superpotential or in the chirality constraint of the Fermi multiplets of the GLSM are not in general mirror symmetric.  In the case of GLSMs based on reflexively plain pairs of polytopes,  there is a possibility of mirror exchange at the level of GLSMs, and this special class of theories certainly deserves further attention, since it may lead to a relatively simple mirror map that includes (0,2) deformations.

Another possibility is that we have perhaps made some unwarranted assumptions about parameters, field redefinitions, or the supersymmetry constraint, which may affect the results.  In this final section of the paper, we will comment on these possibilities.

\subsection{Redundant Parameters in (2,2) GLSMs}
We have argued that the set of deformations of a GLSM can be thought of as the space of holomorphic parameters modulo field redefinitions.  This follows from the assumption that the chiral and twisted chiral F-terms determine the properties of the IR fixed point.  The presence of non-toric and non-polynomial deformations of a (2,2) GLSM already indicates that this is not the entire story:  in general the theory does have additional deformations.  Another caveat that deserves notice is that parameters can be redundant even if they are not absorbable into a D-term by a field redefinition.  While it is easy to see how field redefinitions remove parameters from the chiral superpotential (i.e. complex structure parameters), it is not so clear what happens to the redundant K\"ahler parameters in the twisted chiral superpotential.  We will now explore this point in detail.

Consider an A-twisted GLSM for a C-Y hypersurface $\Mt \subset \Vt$, where the toric fan for $\Vt$ is a maximal projective subdivision  of the toric fan for a toric variety $V$.  As our working example we will take $\Vt$ to be a resolution of $\P^4_{1,1,2,2,6}$, defined by $\Delta^\circ$ with non-zero lattice points given by columns of 
\begin{equation}
\Delta^\circ:
\begin{pmatrix} 
-1 & 1 & 0 & 0 & 0 & ~0 & ~0 \\
-2 & 0 & 1 & 0 & 0 & -1  & ~0 \\
-2 & 0 & 0 & 1 & 0 & -1  & ~0 \\
-6 & 0 & 0 & 0 & 1 & -3  & -1 
\end{pmatrix}.
\end{equation}
This polytope has a unique lattice interior facet point: 
\begin{equation}
\rho_7 =\ff{1}{6} \rho_1 + \ff{1}{6} \rho_2+ \ff{1}{3} \rho_3 + \ff{1}{3} \rho_4.
\end{equation}
The quotient group is $G=(\C^\ast)^3$, with charge matrix
\begin{equation}
Q = 
\begin{pmatrix}
0 & 0 & 1 & 1 & 0 & ~1 & -3 \\
1 & 1 & 0 & 0 & 0 & -2  & ~0 \\
0 & 0 & 0 & 0 & 1 & ~0   & ~1
\end{pmatrix}.
\end{equation}
The anti-canonical class of $\Vt$ is
\begin{equation}
-K = \sum_{\rho} \xi_{\rho} = 2 \eta_3,
\end{equation}
where in the second equality we expressed the $\xi_\rho$ in terms of a basis for $H^{1,1}(\Vt)$:  $\xi_\rho = \sum_a Q^a_\rho \eta_a.$   
In a smooth phase of the GLSM the exceptional set is 
\begin{equation}
F = \{ Z_{1} = Z_{2} = 0\} \cup \{Z_3=Z_4=Z_6=0\} \cup \{Z_5=Z_7=0\}, 
\end{equation}
which leads to the Stanley-Reisner ideal with generating relations
\begin{equation}
\eta_2^2 = 0, \quad \eta_1^2 (\eta_1-2\eta_2) = 0, \quad \eta_3 (\eta_3 - 3\eta_1) = 0.
\end{equation}
The last relation implies that $-K \xi_7 = 0$, which is just the usual statement that the hypersurface $\Mt$ does not intersect the exceptional divisor associated to a lattice interior facet point of $\Delta^\circ$.  It is not hard to show that this holds in general.

The situation is more involved once quantum corrections are taken into account.  To study the quantum cohomology of the $\Mt$ model, we will use the quantum restriction formula of~\cite{Morrison:1994fr}, which relates the quantum cohomology of the $\Mt$ model to that of the $\Vt$ model.  The observables corresponding to $\eta_a \in H^{1,1}(\Vt)$ are the $\sigma_a$, and the relation between the $\Mt$ model and $\Vt$ model A-twisted correlators is
\begin{equation}
\lad \sigma_{a_1} \sigma_{a_2} \sigma_{a_3} \rad  = \la \sigma_{a_1} \sigma_{a_2} \sigma_{a_3} \frac{-K}{1-K} \ra.
\end{equation}
Here the left-hand side denotes a correlator in the $\Mt$ model, while the right-hand side is evaluated in the $\Vt$ model.  Naively, the correlators depend on all three complexified K\"ahler parameters $q_a$, and it is not clear in what sense one of these is redundant.  A little thought suggests the proper point of view:  some combination of the $\sigma_a$ must decouple from the $\Mt$ correlators.   To see how this works, we note that the quantum cohomology relations of the $\Vt$ model include $-K (\sigma_3 - 3\sigma_1) = 2 q_3$.  Using quantum restriction we have
\begin{equation}
\lad \sigma_{a_1}\sigma_{a_2} (\sigma_3-3\sigma_1) \rad =
\la \sigma_{a_1}\sigma_{a_2} \frac{2 q_3}{1-K} \ra = 4 q_3 \lad \sigma_{a_1} \sigma_{a_2} \sigma_3\rad.
\end{equation}
In the last equality we used the ghost number selection rule of the $\Vt$ model.   Since this holds for any $a_1,a_2$, $\sigma_r = 3\sigma_1 + (4q_3-1)\sigma_3$ decouples from the $\Mt$ model correlators.  

Recall that in twisted (2,2) theories there is an intimate relation between local observables and deformations of the action, which in the A-twisted GLSM takes the form
\begin{equation}
\sigma_a \qquad \leftrightarrow \qquad \delta S =  -\frac{1}{4\pi\sqrt{2}} \int d^2 z d\theta^+ d\thetab^- \frac{\delta q_a}{q_a} \Sigma_a.
\end{equation}
Since $\sigma_r$ decouples from the $\Mt$-model correlators, this relation shows that the theory should be independent of the deformation parameter corresponding to $\sigma_r$.  While this is a nice explanation of what redundancy means on the A-model side, it is an example where the redundancy cannot be understood by field redefinitions in the classical action, since a conspiracy of instanton contributions is necessary to produce the decoupling.

It is clear that this discussion will generalize to any $A$-model computation with a redundant $\sigma_a$.  This is easily seen in the mirror B-model point of view.  A redundant divisor $D_\rho$ in the A-model corresponds to a monomial in the mirror polynomial $P^\circ$ associated to an interior facet point $\rho \in \Delta^\circ$, and each such monomial is accompanied by an extra B-model chiral ring relation.  These additional relations can always be used to eliminate the ``redundant'' monomials in favor of the minimal ones,  and the monomial-divisor mirror map will then yield the redundant combinations of the $\sigma_a$.  Although the computations are reasonably simple, in general the results will not have the simple form in the example above.

\subsection{Redundant Deformations of (0,2) GLSMs}
The main tool that makes the treatment of these redundant deformations reasonably straightforward in the (2,2) case is the relation between topological field theory observables and deformations of the action.  Unfortunately, this is precisely the tool we lack in (0,2) theories:  with only (0,2) supersymmetry there is no longer any obvious relation between the $\sigma_a$ and the deformations of the action governed by the $q_a$.  While it seems reasonable to suspect that if the (2,2) theory contains $w$ redundant $q_a$ parameters then the deformed (0,2) theory will also have $w$ redundant parameters that are not eliminated by obvious redefinitions, we cannot prove the assertion.

When compared with the minimal model, the maximal (0,2) model contains additional parameters that involve the fields $Z_{\rho},\Gamma^{\rho}$ where $\rho \in \ri \varphi^\circ$ for some facet $\varphi^\circ$.  A priori, it could be that these additional parameters are also redundant in some fashion that cannot simply be seen by considering the field redefinitions. This is a question that could at least be studied in terms of classical geometry:  does the structure of the deformed bundle actually depend on these parameters?  It would be useful to study this point in detail.

\subsection{The Supersymmetry Constraint}
Another assumption made above is that the supersymmetry constraint imposes the correct number of equations on the set of parameters.  Could it be that the equations are somehow not generic enough, and the number of constraints is not the expected number?  We believe not, and to illustrate the point, we will consider the parameters, redefinitions and supersymmetry for a hypersurface in $(\P^{1})^3$ and its maximal mirror, which have
\begin{equation}
N(M) = 65, \quad\text{and}\quad N(\Mt^\circ) = 59.
\end{equation}
Working with the original theory, the gauge charges are
\begin{equation}
Q = \begin{pmatrix}
-2 & 1 & 1 & 0 & 0 & 0 & 0 \\
-2 & 0 & 0 & 1 & 1 & 0 & 0 \\
-2 & 0 & 0 & 0 & 0 & 1 & 1 
\end{pmatrix}.
\end{equation}
In this case the $E^{a\rho} = \sum_{\rho'} E^{a\rho\rho'} Z_\rho'$, and the field redefinitions consist of: $\GL(2,\C)^3$ transformations of the $Z_{\rho}$ of like charges; $\GL(2,\C)^3$ transformations of the $\Gamma^\rho$ of like charges; the $\GL(3,\C)$ rotations of the $\Sigma_a$; and, finally, rescalings of $\Gamma^0$ and $Z_0$.  Of course, these should be thought of modulo complexified gauge and $\GUL$ transformations.  The latter allows us to forget about the rescaling of $Z_0$.
We fix the remaining redefinitions in several steps.  First, we follow~\cite{Aspinwall:1993rj} and use some of the $Z_\rho$ redefinitions to set to zero the coefficients of monomials in $P$ associated to lattice interior facet points of $\Delta$.  These monomials are just
\begin{eqnarray}
Z_1^2 Z_3 Z_4 Z_5 Z_6, &~~& Z_2^2 Z_3 Z_4 Z_5 Z_6, ~~ Z_1 Z_2 Z_3^2 Z_5 Z_6, \nonumber\\
Z_1 Z_2 Z_4^2 Z_5 Z_6, &~~&  Z_1 Z_2 Z_3 Z_4 Z_5^2, ~~ Z_1 Z_2 Z_3 Z_4 Z_6^2.
\end{eqnarray}
We use the remaining $3$ non-gauge $Z_\rho$ redefinitions and the rescaling of $\Gamma^0$, to fix four of the coefficients in $P$ to $1$:
\begin{equation}
P = Z_1^2 Z_3^2 Z_5^2  + Z_1 Z_2 Z_3^2 Z_5^2 + Z_1 Z_2 Z_4^2 Z_5^2+Z_1 Z_2 Z_3Z_4 Z_5 Z_6+
\text{17 more terms}.
\end{equation}
To fix the $\Gamma^\rho$ rotations, we set some of the coefficients in $J_\rho$ to their $(2,2)$ values, taking, 
\begin{eqnarray}
J_{1} & = & 1\times Z_2 Z_3 Z_4 Z_5 Z_6 + 0 \times Z_1 Z_3 Z_4 Z_5 Z_6 +\ldots, \nonumber\\
J_{2} & = & 0\times Z_2 Z_3 Z_4 Z_5 Z_6 + 1 \times Z_1 Z_3 Z_4 Z_5 Z_6 +\ldots, \nonumber\\
\end{eqnarray}
and similarly for the pairs $J_3,J_4$ and $J_5,J_6$.
Finally, we fix the $\Sigma_a$ rotations by setting $E^{a11}$, $E^{a22}$, and $E^{a33}$ to their (2,2) values.

Having fixed all the redefinitions, we are left with $96$ $\beta$ parameters in $J$, $27$ $\gamma$ parameters in $E^{a\rho}$, and $3$ $\delta$ parameters in $E^{a0}$.  We collect these $126$ parameters into a vector $v$.  These are subject to  the supersymmetry constraint in eqn.~(\ref{eq:constsusy}).  Observe that the constraint takes the general form
$L v =M(v,v) ,$ where $L$ is a linear operator that depends on the $\alpha_m$ and the charges, while $M(w,v)$ is bilinear in its arguments.  This form suggests that we look for a perturbative solution, 
\begin{equation}
v = t v_1 + t^2 v_2 +t^3 v_3 + \cdots.
\end{equation}  To first order in $t$, we just need a solution to the linear system $L v_1  = 0$.  Thus, $v_1$ is in the kernel of $L$.  To second order, we have $L{v_2} = M(v_1,v_1)$, which has a solution provided that $M(v_1,v_1)$ is in the image of $L$.  Similarly, a solution to order $k$ can be extended to order $k+1$ provided that $\sum_{a} M(v_a, v_{k+1-a})$ is in the image of $L$.   

Applying this to the example, we find that $L : \C^{126} \to \C^{81}$, and $\dim \ker L = 45$.  Thus, $L$ is onto, and each element in $\ker L$ extends to a non-linear solution of the supersymmetry constraint.  So, at least in this example, the supersymmetry constraint system of equations is generic enough. It is not hard to repeat the procedure in the mirror model.  We will spare the reader the details, but the final result is the same: the constraint eliminates exactly the expected number of parameters.  Since already in this example $N(M) \neq N(M^\circ)$, the supersymmetry constraint cannot by itself be the cause of the discrepancy.

\appendix

\section{(0,2) Deformations of a Toric GLSM}
The main text has been devoted to the study of (0,2) deformations of linear sigma models that are believed to flow to non-trivial IR fixed points.
It is also interesting to consider (0,2) deformations of massive theories, in which case the deformations are best of thought of in terms
of parameters of the associated half-twisted theory and its topological heterotic ring~\cite{Adams:2005tc}---a deformation of the A-model topological theory.  The simplest case of this sort is a GLSM for a toric variety $V$.  It is, in a sense, a simpler theory, having none of the complications associated to the hypersurface and lacking any J-deformations; however, precisely this simplicity is responsible for a number of subtleties in describing the parameter space.  We will now explore these issues, elaborating and correcting the discussion given in~\cite{McOrist:2007kp,McOrist:2008ji}.

We consider a $V$-model with a (0,2) Lagrangian constructed from the toric data for a compact simplicial toric variety $V$.  This theory depends on the 
K\"ahler parameters $q_a$, as well as the E-parameters in $\cDb_+ \Gamma^\rho = E^\rho(\Sigma,Z)$.  An important difference between the $V$-model and the $M$-models studied above is the anomaly in the $\GUR/\GUL$ symmetries.  The reader will have observed that in the M-model the $\GUR$ implies that the $E^\rho$ must be linear in the $\Sigma$ superfields.  Since this symmetry is now broken by quantum effects, it is not clear that the $E^\rho$ must remain linear in the $\Sigma$s.  We will not consider E-deformations that are non-linear  in the $\Sigma$, choosing to preserve the classical $\GUR$ symmetry.

\subsection{Parameters and Redefinitions}
To discuss the V-model deformations and field redefinititions, it is convenient to split the matter fields according to the gauge charges~\cite{McOrist:2007kp}:
\begin{equation}
\{ (\Gamma^\rho,Z_\rho) ~|~ \rho \in \Sigma_V(1) \} \to \cup_{\alpha=1}^k \{ (\Gamma^{\rho_\alpha}_{(\alpha)},Z_{(\alpha),\rho_{\alpha}})~|~ \rho_\alpha \in I_{\alpha}\},
\end{equation}
such that any two fields belonging to the same set $I_{\alpha}$ have the same gauge charges $Q^a_{(\alpha)}$.\footnote{If $\Sigma_V(1)$ does not generate the lattice $N$, so that the gauge group has an additional factor of a finite discrete group $H$, the fields should be further split up according 
to representations of $H$.}   The $n_\alpha = |I_{\alpha}|$ satisfy $\sum_{\alpha} n_\alpha = n$.  With this notation the most general $E^\rho$ takes the form
\begin{equation}
\cDb_+ \Gamma_{(\alpha)} = \sum_{a=1}^r \Sigma_a \{ M^a_{(\alpha)} Z_{(\alpha)} + N^a_{(\alpha)}(Z)\},
\end{equation}
where $M^a_{(\alpha)}$ is a field-independent $n_\alpha \times n_\alpha$ matrix, and $N^a_{(\alpha)}$ is a column vector with entries non-linear in the $Z_{\rho}$.  The total number of parameters in the $M^a_{(\alpha)}$ and $N^a_{(\alpha)}$ is $ r \times \dim\Autt V$.

We will first $V$-models that do not allow non-linear terms in the $E^\rho$.  The field redefinitions consistent with gauge invariance are 
\begin{equation}
\Gamma_{(\alpha)} \mapsto V_{\alpha} \Gamma_{\alpha}, \quad Z_{\alpha} \mapsto U_{\alpha} Z_{\alpha}, \quad \Sigma_a \mapsto \sum_{b=1}^r\Sigma_b G^b_a,
\end{equation}
where $V_{(\alpha)}$ and $U_{(\alpha)}$ are field-independent $\GL(n_\alpha,\C)$ matrices, and $G \in \GL(r,\C)$.
These act on the $M^{a}_{(\alpha)}$ by 
\begin{equation}
M^{a}_{(\alpha)} \mapsto \sum_{b=1}^r G^a_b V_{(\alpha)}^{-1} M^b_{(\alpha)} U_{(\alpha)}.
\end{equation}
The redefinitions depend  on $2 \times \dim\Autt V +r^2$ parameters; however, unlike in the M-model, not all of these act effectively.  This point was missed in an earlier analysis by two of us~\cite{McOrist:2008ji}.  First, the $E^{a\rho}$ set to their (2,2) values are left invariant by the (2,2) redefinitions with $U_{(\alpha)} = V_{(\alpha)}$.  These redefinitions are just the complexification of the global symmetry group $\prod_\alpha \GU(n_\alpha)$.  Second, the (2,2) theory has a classical global $\GUL$ symmetry, which in a product theory with a target space $V_1 \times \cdots \times V_s$ is enhanced to $\left[\GU(1)\right]^s$.  Its complexification also fixes the (2,2) locus.  
Putting this together, we expect that
\begin{eqnarray}
\label{eq:Vsimp}
N(V) &=& r + r \sum_{\alpha} n_\alpha^2 - \left[ \sum_\alpha n_\alpha^2 + r^2 -s \right] \nonumber\\
~ & = & ~ r + (r-1) \sum_{\alpha} n_\alpha^2 -r^2 + s.
\end{eqnarray}

It is worthwhile to verify this by a more explicit analysis. Expanding to first order around the (2,2) locus, it is clear that the redefinitions will
fix the (2,2) locus provided that
\begin{equation}
\sum_{b=1}^r \delta G^a_b Q^b_{(\alpha)} {\mathbbm  1}_{(\alpha)} + (\delta U_{(\alpha)} - \delta V_{(\alpha)} )Q^a_{(\alpha)} = 0~~\text{for all}~~a,\alpha.
\end{equation}
This equation has solutions if and only if
\begin{equation}
\delta V_{(\alpha)} =  \delta U_{(\alpha)}-\delta w_{(\alpha)} {\mathbbm  1}_{(\alpha)}
\end{equation}
for some set of $k$ constants $\delta w_{(\alpha)}$ that satisfy
\begin{equation}
\label{eq:deltaG}
\sum_{b=1}^r \delta G^a_b Q^b_{(\alpha)}  = \delta w_{(\alpha)} Q^a_{(\alpha)}.
\end{equation}
When considered as an $r \times k$ matrix, $Q^a_{(\alpha)}$ must have rank $r$.  Suppose we arrange the $Q^a_{(\alpha)}$ so that 
$Q^a_{(1)},\ldots,Q^a_{(r)}$  are linearly independent, while the rest can be expressed as
\begin{equation}
Q^a_{(\mu)} = \sum_{b \in R_\mu} \tau^b_\mu Q^a_{(b)},
\end{equation}
where $R_\mu \subset \{1,\ldots, r\}$ such that $\tau^b_\mu \neq 0$.  Requiring that eqn.~(\ref{eq:deltaG}) holds for $\alpha = 1, \ldots, r$ determines $\delta G^a_b$ in terms of the charges and the $\delta w_{(a)}$, while the remaining $k-r$ equations require $\delta w_{(a)} = \delta w_{(\mu)}$ for each $a \in R_\mu$.  The non-effective redefinitions are parametrized by $\delta U_{(\alpha)}$ and the $\delta w_{(\alpha)}$ subject to the constraint just described.  We conclude that a V-model without non-linear E-parameters has at the linearized level
\begin{equation}
 N(V) = \underbrace{r}_{\text{K\"ahler}} + \underbrace{ r \sum_{\alpha} n_\alpha^2}_{E} 
 -\underbrace{\left[ \sum_\alpha n_\alpha^2 + r^2 - \#\delta w\right]}_{\text{effective redefinitions}}. 
\end{equation}
For each $\delta w$ parameter in addition to the one associated to the overall $\GUL$ symmetry, there must be a subset of matter fields charged under a subgroup $G'$ of the gauge group $G$ and neutral under $G/G'$.  Thus, the (2,2) GLSM is actually a product theory with target space a product of toric varieties, and each factor contains an independent $\GUL$ symmetry. A $\delta w$ parameter corresponds to a linearization of the complexified $\GUL$ rotation for the factor.  Thus, we reproduce the counting advertised above in eqn.~\ref{eq:Vsimp}.

\subsection{Non-Linear E-deformations}
Having described the situation where the $E^\rho$ are linear in the $Z_\rho$, we now turn to the more general situation where $N^a_{(\alpha)}$ are non-zero, and there are additional possible redefinitions
\begin{equation}
\Gamma_{(\alpha)} \mapsto V_{(\alpha)} \Gamma_{(\alpha)} + \sum_\beta V'_{(\alpha\beta)}(Z) \Gamma_{(\beta)}, \quad Z_{(\alpha)} \mapsto U_{(\alpha)} Z_{(\alpha)} + U'_{(\alpha)}(Z).
\end{equation}
The (2,2) locus is fixed by transformations with $V_{(\alpha)} = U_{(\alpha)}$ and 
\begin{equation}
(V'_{(\alpha\beta)})_{\rho_\alpha \rho_{\beta}}  =  \frac{\p (U'_{(\alpha)})_{\rho_\alpha}}{\p Z_{\rho_\beta}},
\end{equation}
as well as any R-symmetries.  Thus, we expect that in any V-model GLSM with target space $V = V_1 \times \cdots \times V_s$, the number of parameters is given by
\begin{equation}
N(V) = r + (r-1) \dim \Autt V -  r^2 +s .
\end{equation}

\subsection{Do non-linear parameters matter?}
There is one more caveat that we should make.  The preceding counting assumes that the non-linear E-parameters do actually affect the A/2 half-twisted correlators.  However, at least as far as smooth theories are concerned, this is not entirely clear.  For instance, when we compute the correlators by summing over $\sigma$-vacua in a non-geometric phase, large $\sigma$-vevs for a smooth $V$-model suppress any contributions from interactions involving the $N^a_{(\alpha)}(Z)$.  Moreover, there are certainly models where the non-linear parameters do not affect some crude features of the theory, such as the classical discriminant locus.

\begin{figure}[h!]
\[
\begin{xy} <1.2mm,0mm>:
  (0,0)*{} ="0", (15,0)*{}="pX", (-15,0)*{}="mX",
  (0,15)*{} = "pY", (0,-15)*{} ="mY", (15,20)*{\Delta^\circ \subset N_\R},
  (0,5)*{\bullet}="v1", (5,-5)*{\bullet}="v2" , (0,0)*{\bullet} , (-5,-5)*{\bullet}="v3", (0,-5)*{\bullet}="v4",
  (0,8)*{3}, (-5,-8)*{1},(0,-8)*{4}, (5,-8)*{2},
\ar@{-}|{} "mX"; "pX" 
\ar@{-}|{} "mY"; "pY"
\ar@{-}|{} "v1"; "v2"
\ar@{-}|{} "v2"; "v3"
\ar@{-}|{}"v3"; "v1"
\end{xy}
\]
\caption{The polytope for ${\mathbbm F}_2$.}
\label{fig:tents}
\end{figure}
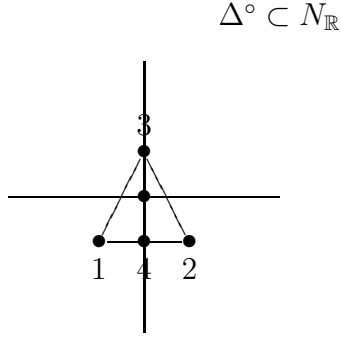

Let us elaborate on the last point in a simple example.  We take $V$ to be the Hirzebruch surface ${\mathbbm F}_2$, with polytope given in figure~\ref{fig:tents}.
The charges in this case are
\begin{equation}
Q = \begin{pmatrix} 1 & 1& 1& -1 \\ 1 & 1 & 0 & -2 \end{pmatrix},
\end{equation}
and the most general $E^{a\rho}$ is given by
\begin{equation}
E^{a\rho} = \begin{pmatrix} 
\alpha_1 Z_1 + \beta_1 Z_2 & \delta_1 Z_1 + \ep_1 Z_2& c_1 Z_3 + f Z_4 & d_1 Z_4 \nonumber \\
\alpha_2 Z_1 + \beta_2 Z_2 & \delta_2 Z_1 + \ep_2 Z_2& c_2 Z_3 + g Z_4 & d_2 Z_4
\end{pmatrix},
\end{equation}
where $f$ and $g$ are degree two homogeneous functions of $Z_1,Z_2$.   In the smooth phase of the GLSM the
excluded set is $F_\Sigma = \{Z_1 = Z_2 = 0\} \cup \{Z_3 = Z_4 = 0\}$.  In general, the matrix $E$ defines a sheaf $\cE$
over the toric variety $V$ via the exact sequence
\begin{equation}
\xymatrix{ 0\ar[r] &\cO^{r}\ar[r]^-{E} &\oplus_\rho\cO(D_\rho) \ar[r] & \cE\ar[r] & 0 }.
\end{equation}
 Away from the discriminant locus $\cE$ is a rank $2$ bundle, and with parameters set to 
their (2,2) values, the bundle is just $T_V$.  To compute the classical discriminant locus, we can study the conditions 
under which for some $Z \not\in F_\Sigma$  the rank of $E^{a\rho}$ becomes less than $2$.  This is a straightforward
computation, and the result is that the rank drops to $1$ at some point in $V$ if and only if the parameters satisfy $D_1 =0$
or $D_2 = 0$, with
\begin{eqnarray}
D_1 &=& c_2^2 \Delta_{11} -2 c_1 c_2 \Delta_{12} + c_1^2 \Delta_{22}, \nonumber\\
D_2 &=& d_2^2 \Delta_{11} -2 d_1 d_2 \Delta_{12} + d_1^2 \Delta_{22},
\end{eqnarray}
where
\begin{eqnarray}
\Delta_{11} & = & \alpha_1 \ep_1 -\beta_1 \delta_1, \nonumber\\
2\Delta_{12} & = & \alpha_1 \ep_2 -\beta_1 \delta_2 + \alpha_2 \ep_1 -\beta_2 \delta_1, \nonumber\\
\Delta_{22} & = & \alpha_2 \ep_2 - \beta_2 \delta_2.
\end{eqnarray}
As promised, $D_1$ and $D_2$ do not depend on the parameters in $f$ and $g$. The non-linear
parameters do affect the set of points in $V$ where the rank jumps.  Namely, if $D_2 = 0$ and $c_1 d_2 -c_2 d_1 = 0$ (these
two conditions also imply $D_1 = 0$), if the non-linear parameters are non-zero the rank continues to jump at points, while
if they are set to zero, the rank jumps on a curve.

While admittedly crude, this suggests that the non-linear deformations do not affect the properties of the V-model away from
the singularities.  This should be contrasted with the $M$-model, where the interplay between the E- and J- parameters
via the supersymmetry constraint should give the non-linear E-parameters a more significant role to play.


\providecommand{\href}[2]{#2}\begingroup\raggedright\endgroup

\end{document}